\documentclass[a4paper,11pt]{article}
\pdfoutput=1 
\usepackage{jcappub} 
\usepackage[T1]{fontenc} 
\usepackage{bm}
\usepackage{booktabs}
\usepackage{adjustbox}
\usepackage{graphicx}
\usepackage{subcaption}
\usepackage{orcidlink}
\usepackage{mathrsfs}
\PassOptionsToPackage{dvipsnames}{xcolor}
\usepackage[normalem]{ulem}

\title{\boldmath The Impact of Spectroscopic Redshift Errors on Cosmological Measurements}

\author[a]{Shengyu He\orcidlink{0009-0007-4971-8231},}
\author[a,b]{Jiaxi Yu\orcidlink{0009-0001-7217-8006},}
\author[a]{Antoine Rocher,\orcidlink{0000-0003-4349-6424}}
\author[a,c]{Daniel Forero-S\'anchez,}
\author[a]{Jean-Paul Kneib,}
\author[d]{Cheng Zhao,}
\author[e]{Etienne Burtin,}
\author[f,g]{Jiamin Hou}

\affiliation[a]{Institute of Physics, Laboratory of Astrophysics, \'Ecole Polytechnique F\'ed\'erale de Lausanne (EPFL), Observatoire de Sauverny, CH-1290 Versoix, Switzerland}

\affiliation[b]{Kavli Institute for the Physics and Mathematics of the Universe (WPI), The University of Tokyo Institutes for Advanced Study (UTIAS), The University of Tokyo, Chiba 277-8583, Japan}

\affiliation[c]{Institut de Ciències del Cosmos (ICCUB), Universitat de Barcelona (UB), c. Martí i Franquès, 1, 08028 Barcelona, Spain.}

\affiliation[d]{Department of Astronomy, Tsinghua University, Beijing 100084, China}

\affiliation[e]{IRFU, CEA, Universit\'e Paris-Saclay, F-91191 Gif-sur-Yvette, France}

\affiliation[f]{Institute of Astronomy, University of Cambridge, Madingley Rd, Cambridge CB3 0HA, UK}

\affiliation[g]{Kavli Institute for Cosmology Cambridge, Madingley Road, Cambridge CB3 0HA, UK}

\emailAdd{shengyu.he@epfl.ch, jiaxi.yu@ipmu.jp, antoine.rocher@epfl.ch}

\newcommand{\h}{h}
\newcommand{\Om}{\Omega_\mathrm{m}}
\newcommand{\Ob}{\Omega_\mathrm{b}}

\newcommand{\ns}{n_s}
\newcommand{\aiso}{\alpha_\mathrm{iso}}
\newcommand{\aap}{\alpha_\mathrm{ap}}
\newcommand{\apar}{\alpha_\parallel}
\newcommand{\aper}{\alpha_\perp}

\newcommand{\logA}{\ln(10^{10} A_{s})}
\newcommand{\sumnu}{\sum m_\nu}

\newcommand{\vpx}{v_{{\rm{pec}},x}}
\newcommand{\dv}{\Delta v}
\newcommand{\dve}{\Delta v_{\rm{error}}}

\newcommand{\dvs}{\Delta v_{\rm{smear}}}
\newcommand{\sigmadv}{\sigma_{\Delta v}}
\newcommand{\wdv}{\mathrm{w}_{\Delta v}}
\newcommand{\sigmaz}{\sigma_{\rm{zerr}}}
\newcommand{\wz}{\mathrm{w}_{\rm{zerr}}}
\newcommand{\fc}{f_c}
\newcommand{\dvc}{\Delta v_{\rm{c}}}
\newcommand{\dvch}{\Delta \hat{v}_{\rm{c}}}
\newcommand{\dz}{\Delta z}

\newcommand{\LCDM}{\Lambda \rm{CDM}}

\newcommand{\wCDM}{w_0w_a\rm{CDM}}

\newcommand{\QUIJOTE}{\textsc{Quijote}~}
\newcommand{\FOLPS}{\textsc{FOLPS$\nu$}}
\newcommand{\desilike}{\texttt{desilike}}
\newcommand{\iminuit}{\texttt{iminuit}}

\newcommand{\getdist}{\texttt{GetDist}}
\newcommand{\emcee}{\texttt{emcee}}
\newcommand{\EFT}{\texttt{EFT}}
\newcommand{\EFTfreefc}{\texttt{EFT}+{\rm{free}}\,f_c}
\newcommand{\EFTfixedfc}{\texttt{EFT}+{\rm{fixed}}\,f_c}

\newcommand{\hmpc}{h \, \rm{Mpc}^{-1}}
\newcommand{\kmps}{\rm{km} \cdot s^{-1}}
\newcommand{\hgpc}{h^{-3} \rm{Gpc}^3}

\usepackage[nameinlink,noabbrev]{cleveref}
\crefname{equation}{Eq.}{Eqs.}

\crefname{figure}{Figure}{Figures}
\crefname{table}{Table}{Tables}
\crefname{appendix}{Appendix}{Appendices}
\Crefname{figure}{Figure}{Figures}
\Crefname{equation}{Equation}{Equations}
\Crefname{section}{Section}{Sections}
\Crefname{table}{Table}{Tables}

\abstract{Spectroscopic redshift errors, including redshift uncertainty and catastrophic failures, can bias cosmological measurements from galaxy clustering at sub-percent level. In this work, we investigate their impact on the full-shape analysis using contaminated mock catalogs. We find that redshift uncertainty introduces a scale-dependent damping effect on the power spectrum, which is absorbed by counterterms in clustering model, keeping parameter biases below $5\%$. Catastrophic failures suppress the power spectrum amplitude by an approximately constant factor, which scales with the catastrophic rate $f_c$. While this effect is negligible for DESI galaxy populations ($f_c=1\%$), the slitless-like errors, combining redshift uncertainty with $f_c=5\%$ catastrophics, introduce significant biases in cosmological constraints. In this case, we find $6\%$ to $16\%$ shifts ($\sim2.2\sigma$ level) in estimating the fractional growth rate $df\equiv f/f^{\rm{fid}}$ and the log primordial amplitude $\ln(10^{10} A_{s})$. Applying the correction factor $(1-f_c)^2$ on the galaxy power spectrum mitigates the bias but weakens the parameter constraints due to new degeneracies. Alternatively, fixing $f_c$ to its expected value restores the constraining power with a modest bias of $1.0\sigma$. Our results indicate that for space-based slitless surveys such as \textit{Euclid}, at minimum accurate estimation of $f_c$ and its incorporation into the clustering model are necessary to get unbiased cosmological inference. Extending to evolving dark energy and massive neutrino cosmologies, redshift errors do not bias the dark energy properties parametrized by $w_0$ and $w_a$, but can degrade constraints on the summed neutrino mass $\sum m_\nu$ by up to 80\% in the worst case.} 

\begin{document}
\maketitle
\flushbottom

\newpage

\section{Introduction}\label{sec:intro}

The large-scale structure (LSS) observed by galaxy redshift surveys has played an important role in advancing our understanding of fundamental physics. Clustering features, such as baryon acoustic oscillations (BAO; \cite{eisenstein_baryonic_1998, eisenstein_detection_2005, weinberg_observational_2013, alam_clustering_2017, desi_collaboration_desi_2025-4}), and the full-shape power spectrum \cite{ivanov_cosmological_2020, damico_cosmological_2020, desi_collaboration_desi_2025-1, desi_collaboration_desi_2025-3}, have been used to constrain some key quantities in cosmology, including dark energy equation of state $w$, the matter density $\Om$, the structure growth rate $f\sigma_8$, neutrino mass $\sumnu$ and possible deviations from general relativity. Current leading experiments, the Dark Energy Spectroscopic Instrument\footnote[1]{DESI, \href{https://www.desi.lbl.gov}{desi.lbl.gov}} (DESI; \cite{desi_collaboration_desi_2016, desi_collaboration_desi_2016-1}), and the ongoing space mission \textit{Euclid}\footnote[2]{\textit{Euclid}, \href{https://www.cosmos.esa.int/web/euclid}{cosmos.esa.int/web/euclid}} \cite{laureijs_euclid_2011, euclid_collaboration_euclid_2025-2} are mapping the three-dimensional Universe by measuring redshifts for tens to hundreds of millions galaxies. These surveys will cover over more than 15,000 $\rm deg^2$ of the sky, and trace the LSS back to redshift $\sim3.5$. This large probing volume allows the clustering signal to be measured with minimal cosmic variance, leading to sub-percent precision in cosmological measurements. Given this level of precision, detailed studies on systematic errors are critical to ensure the robustness and reliability of cosmological constraints.

Among various observational systematics, we focus on the impact of spectroscopic redshift errors. They persist in data with secure redshifts, where redshift calibration and algorithmic processing have already been applied~\cite{blake_wigglez_2010, bolton_spectral_2012, guy_spectroscopic_2023, euclid_collaboration_euclid_2025-1}. There are two primary types of errors: redshift uncertainty and redshift catastrophic failures. Redshift uncertainty refers to minor shifts ($\dz \sim10^{-4}$) in all redshift measurements, primarily caused by instrumental noise and the intrinsic width of spectral features. In contrast, redshift catastrophic failures result from the misidentification of spectral lines, leading to severely incorrect redshifts ($\dz \sim10^{-2}$) in a small fraction of data. These systematic errors can distort the observed clustering signal and potentially bias the cosmological inference. While previous studies have examined their individual impacts \cite{blake_wigglez_2011, zarrouk_clustering_2018, hou_completed_2020, yu_elg_2025, krolewski_impact_2025, euclid_collaboration_euclid_2025-5}, a comprehensive analysis that considers all these effects remains unexplored. This work aims to bridge that gap and explore their combined influence on cosmological inference from full-shape clustering analysis.

The spectroscopic redshift errors can be quantified using repeated redshift observations and incorporated into realistic galaxy mocks (e.g., \cite{smith_completed_2020, yu_model_2022, yu_desi_2023, yu_elg_2025}). By comparing analyses on clean mocks with those contaminated by redshift errors, we can assess their impact on clustering statistics and cosmological constraints. Improved facilities and increased statistics in DESI \cite{desi_collaboration_data_2025, yu_elg_2025} provide a better way to model the redshift errors and study their impact on cosmological analyses. Additionally, space-based surveys such as \textit{Euclid} use slitless spectroscopy at near-infrared to measure redshifts \cite{euclid_collaboration_euclid_2025-4, euclid_collaboration_euclid_2025-3}. This method requires no pre-selection of targets and is particularly effective for probing high-redshift galaxies. However, it is also more susceptible to redshift errors due to the lower spectrum resolution and the potential overlap of different spectra. These challenges have been addressed in the context of redshift interlopers for \textit{Euclid} \cite{euclid_collaboration_euclid_2025-5}. In this work, we use hypothetical combined redshift errors to represent the conditions expected in slitless surveys. This approach provides a description of the related systematics and offers practical guidance for interpreting and mitigating their impact in cosmological analyzes for spectroscopic surveys.

The outline of the paper is as follows: \cref{sec:data spectroscopic survey and redshift errors} introduces two galaxy redshift surveys, DESI and \textit{Euclid}, and the spectroscopic redshift errors. \cref{sec:data contaminated mocks} describes how we construct our contaminated mock catalogs. \cref{sec:methods} presents the methodologies used to derive cosmological inferences from the mock data. In \cref{sec:results}, we present the impact of redshift errors on clustering statistics and how these propagate into the cosmological parameter constraints, including extensions $\wCDM$ and massive neutrino. Finally, we summarize our findings and conclude in \cref{sec:conclusions}.

\section{Spectroscopic Survey and Redshift Errors}
\label{sec:data spectroscopic survey and redshift errors}

Spectroscopic surveys measure the redshifts of objects with high precision by analyzing their spectral features, such as emission and absorption lines. Two massive ongoing spectroscopic surveys, DESI and \textit{Euclid}, are measuring galaxy clustering with unprecedented accuracy. In its first three years, the ground-based survey DESI has observed over 14 million galaxies~\cite{desi_collaboration_desi_2025-4} and plans to extend its initial plan to 60M galaxies across $18{,}000\,\rm{deg}^2$. Its dark matter tracers include Bright Galaxies (BGS) at $0.1<z<0.4$ \cite{hahn_desi_2023}, Luminous Red Galaxies (LRG) at $0.4<z<1.1$~\cite{zhou_target_2023}, Emission Line Galaxies (ELG, mainly [O\,\textsc{ii}] emitters) at $0.6<z<1.6$~\cite{raichoor_target_2023}, and Quasi-Stellar Objects (QSO) at $0.8<z<3.5$~\cite{chaussidon_target_2023}. In parallel, the \textit{Euclid} mission is conducting slitless spectroscopic survey from space and aim to observe tens of millions of galaxies over $15{,}000\,\rm{deg}^2$. Its primary targets are [$\rm{H \, \alpha}$] emitters in the redshift range $0.9<z<1.8$ \cite{euclid_collaboration_euclid_2025-2}. The \textit{Euclid} Quick Data Release 1 (QR1) is already publicly available~\cite{euclid_collaboration_euclid_2025}. The large cosmic volumes probed by DESI and \textit{Euclid} enable sub-percent precision in cosmological measurements and require a rigorous understanding of systematic errors, particularly those related to redshift measurements.

In this work, the spectroscopic systematic redshift errors are referred to as \textit{redshift errors}. The redshift errors exist in both ground-based and space-based galaxy redshift surveys. These include redshift uncertainty, the small statistical fluctuations in redshift measurements, and redshift catastrophic failures, the large errors from the true values. Both types of error exists in secure redshift measurements, i.e., those that have been calibrated and successfully passed the redshift determination pipeline. Consequently, these hidden errors can distort the observed clustering pattern and potentially bias our cosmological analyses. Since both redshift uncertainty and catastrophic failures can be quantified from repeated redshift observations, e.g. $dv_{ij}= c(z_i-z_j )/(1+z_i)$, it is practical to analyze them together. This combined treatment is also relevant for space missions such as \textit{Euclid}, which apply slitless spectroscopy to measure redshifts and are more susceptible to both types of contamination. In the following, we will describe how redshift uncertainty and redshift catastrophic failures are characterized and modeled, and what are the redshift errors expected in slitless spectroscopic surveys.


\subsection{Redshift uncertainty}\label{sec:data redshift uncertainty}

Redshift uncertainty is an inevitable systematic error in spectroscopic redshift measurements. It arises because the spectral lines used to determine redshifts have specific widths rather than being ideal delta functions. These widths can result from a combination of physical (e.g., galaxy thermal motions, turbulence, gravitational broadening) and observational (e.g, instrumental resolution, observing condition) effects. The values of redshift uncertainty are reported for different types of targets within different redshift surveys. For example, in eBOSS galaxy samples, the redshift accuracies, characterized by the root mean square (RMS) of $dv_{ij}$, are about 65.6, 20 and 300$\,\kmps$ for LRG, ELG and QSO~\cite{ross_completed_2020, raichoor_completed_2020, lyke_sloan_2020}. From DESI Visual Inspection, the redshift precisions, defined as the median absolute deviation (MAD) values of $dv_{ij}$, reaches about 10, 10, 40 and 100$\,\kmps$ for BGS, ELG, LRG and QSO respectively~\cite{lan_desi_2023, alexander_desi_2023}. 

The effect of redshift uncertainty is equivalent to randomly smearing the peculiar velocity of an observed object along the line-of-sight. This smearing velocity, $\dvs$, defined as the same as $dv_{ij}$, is empirically quantified using redshift difference. The distribution of observed $\dvs$ can be statistically modeled. For galaxies, this distribution is typically Gaussian $\mathcal{N}(\mu ,\, \sigmadv)$, with $\sigmadv$ characterizing the redshift uncertainty dispersion \cite{ross_completed_2020, yu_model_2022, yu_desi_2023}: 
\begin{equation}\label{eq:gaussian uncertainty}
     \mathcal{N}(\dvs \, |\, \mu ,\, \sigmadv)= \frac{1}{\sqrt{2\pi \sigmadv}} \exp\left(-\frac{(\dvs  -\mu)^2}{2\sigmadv}\right)
\end{equation}
For QSO, the distribution of $\dvs$ usually exhibits extended wide tails. In such case, a multiple-Gaussian model~\cite{lyke_sloan_2020, chaussidon_target_2023} or a Lorentzian distribution~\cite{yu_desi_2023} provide a better fit to repeat observations. In our work, we use the Lorentzian profile, $\mathcal{L}(\,p ,\, \wdv)$, to describe the QSO redshift uncertainty with extended tail:
\begin{equation}\label{eq:lorentzian uncertainty}
     \mathcal{L}(\dvs  \, |\, p ,\, \wdv) = \frac{1}{1+((\dvs -p)/\wdv)^2}
\end{equation}
Here, $\wdv$ captures the width of the distribution, thus characterizing the redshift uncertainty in the Lorentzian profile. To assess how the presence of long-tailed uncertainty distributions may impact clustering and cosmological measurements, we adopt both Gaussian and Lorentzian profiles to model QSO redshift uncertainty. A comparison between these profiles is shown in the left panel of \cref{fig:hist_model_redshift_errors}. 

\subsection{Redshift catastrophic failures}\label{sec:data redshift catastrophics}

Redshift catastrophic failures (or redshift catastrophics) arise in a small fraction of redshift measurements, typically identified through significant redshift differences in repeated observations. These errors can result from misidentification of sky residuals or confusion between spectral lines of different galaxies~\cite{blake_wigglez_2010, yu_elg_2025, euclid_collaboration_euclid_2025-1}. Faint or ambiguous spectral features can also lead to catastrophically incorrect redshifts. The catastrophic contamination rate, $\fc$, and the threshold for identifying it vary in different surveys. For example, the WiggleZ survey reports 5\% $\fc$ of redshift blunders with $|\Delta z| > 400 \, \kmps$ (larger than $4\sigmadv$)~\cite{blake_wigglez_2010}. In eBOSS, the reported $\fc$ is less than 1\% for LRGs and ELGs ($|\Delta z| > 1000\,\kmps$) and 1.6\% for quasars ($|\Delta z| > 3000\,\kmps$)~\cite{raichoor_completed_2020, ross_completed_2020, lyke_sloan_2020}. For DESI DR1, using the same $|\Delta z|$ thresholds as in eBOSS, the $\fc$ is estimated to be less than 1\% for all tracers at $z<2.1$~\cite{desi_collaboration_data_2025}. 

In the study of DESI DR1 repeated ELG observations, redshift catastrophics are classified into two main categories: sky confusion at $z=1.32$ and random catastrophics \cite{yu_elg_2025}. Sky confusion, responsible for 24\% of these errors, arises from misidentifying residuals of sky emissions at around $8625\,\text{\AA}$ and can be approximated as a Gaussian distribution centered at the corresponding sky line positions. The remaining 76\% are random catastrophic failures that encompass various error patterns. These can be modeled using a log-normal profile defined as follows:
\begin{equation}
    \mathcal{LN}(\dvch  |\, \mu_{\text{ran}}, \sigma_{\text{ran}}^2, \hat{v}_0) = \frac{1}{\sqrt{2\pi} \sigma_{\text{ran}}(-\dvch+\hat{v}_0) } \, \exp\left\{-\frac{\left[\ln(-\dvch+\hat{v}_0) \!-\! \mu_{\text{ran}}\right]^2}{2\sigma_{\text{ran}}^2}\right\}
    \label{eq:catastrophics profile}
\end{equation}
This function models the absolute log-normal distribution of redshift catastrophics ($\dvch = \log_{10}(|dv_{ij}|)$) with parameters $\mu_{\text{ran}}$, $\sigma_{\text{ran}}^2$ and $\hat{v}_0$. In our study, we focus on the random pattern while varying the contamination rate $\fc$ to generalize catastrophic patterns for different surveys. We note that including sky confusion would not affect significantly the results~\cite{yu_elg_2025}. The distributions of catastrophics for $\fc=1\%$ and $5\%$ are shown in the right panel of \cref{fig:hist_model_redshift_errors}, with the magnitude of $\dvc$ varying in $10^3$–$10^6 \, \kmps$.

\subsection{Combined errors for Slitless redshift surveys}\label{sec:data slitless survey}

Space missions such as \textit{Euclid}~\cite{euclid_collaboration_euclid_2025-2}, Roman~\cite{spergel_wide-field_2015} and CSST (scheduled for launch in 2027;~\cite{csst_collaboration_introduction_2025}) employ slitless spectroscopy to map the 3D distribution of galaxies over wide areas of sky. Early dataset from \textit{Euclid} QR1 is now publicly available~\cite{euclid_collaboration_euclid_2025} and provides initial evaluations of the performance of space spectroscopic redshift measurement. Unlike traditional ground slit-based spectroscopy that use fibers to capture individual targets, slitless spectroscopy disperses the light from all sources in the field simultaneously, enabling efficient and continuous observations without the need for target pre-selection. This approach allows for high sky coverage and dense sampling of galaxies in the redshift range relevant to LSS studies.

However, the slitless design magnifies the redshift errors. \textit{Euclid} spectroscopic redshifts are derived from spectra provided by the Near Infrared Spectro-Photometer (NISP;~\cite{euclid_collaboration_euclid_2025-4}), which operates with a resolution of $R \simeq 500$~\cite{euclid_collaboration_euclid_2025-3}. In comparison, the ground-based DESI survey uses instruments with resolution $R\sim 2000$ to $5500$ \cite{desi_collaboration_desi_2016}. The lower spectral resolution and slitless spectroscopy pipeline lead to broader spectral lines and consequently increased redshift uncertainty. Additionally, the lack of spatial isolation between sources causes catastrophic redshift errors~\cite{euclid_collaboration_euclid_2025-1, euclid_collaboration_euclid_2025-3}, typically through cross-contamination and self-contamination. Cross-contamination arises from overlapping spectrograms of neighboring sources, while self-contamination results from the degeneracy between spatial and spectral information along the dispersion direction. These combined effects lead to a higher rate of redshift catastrophic failures in the sample~\cite{euclid_collaboration_euclid_2025-5}.

To address these challenges, we adopt a hypothetical redshift error that combines both redshift uncertainty and catastrophics to simulate the expected observational systematics in slitless surveys. This model assumes a larger uncertainty than those typical of ELG (i.e., LRG-like, $\sigma_{\dv} = 85.7\, \kmps$) and a higher rate of catastrophic failures ($\fc=5\%$ with $|\dv|>1000\,\kmps$). For reference, \cite{euclid_collaboration_euclid_2025-1} reports that the scatter of redshift difference between \textit{Euclid} and DESI is $\sigma_{dz} = 0.8\times10^{-3} \approx 240 \, \rm{km/s}$, and the overall success rate indicates that a catastrophic rate $f_c \gtrsim 10\%$ is expected. Therefore, our hypothetical scenario remains relatively optimistic compared to the current expectations from slitless survey.

\section{Contaminated Mocks}
\label{sec:data contaminated mocks}

\begin{figure}
    \centering
    \includegraphics[width=\linewidth]{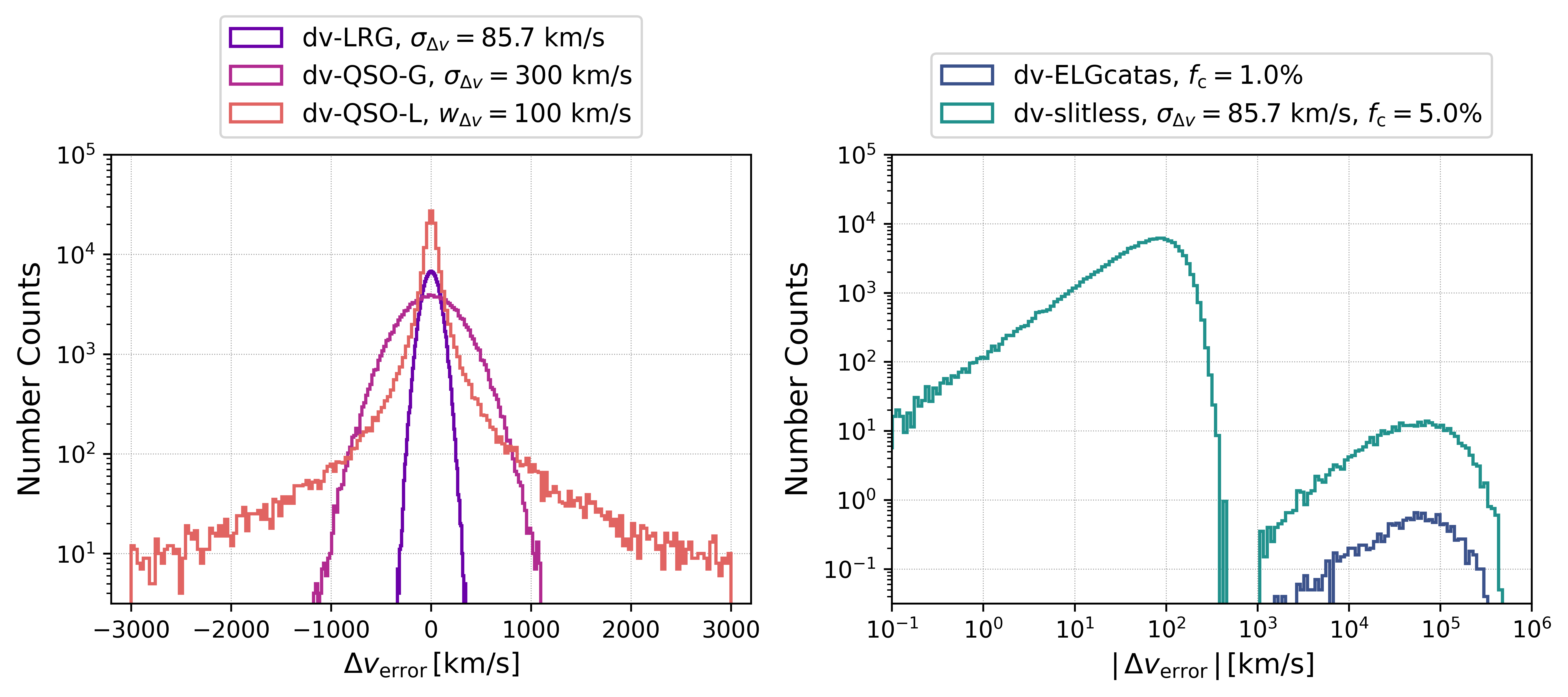}
    \caption{Comparisons of redshift error ($\dve$) distributions. \textit{Left}: Typical LRG-like and QSO-like redshift uncertainties, modeled by Gaussian smearing with $\sigmadv$ or Lorentzian smearing with $\wdv$. \textit{Right}: ELG-like catastrophics with $\fc=1\%$ and hypothetical slitless-like errors with LRG-like redshift uncertainty (left part) and $\fc=5\%$ catastrophics (right part). These distributions are used to generate the velocity dispersion caused by redshift errors.}
    \label{fig:hist_model_redshift_errors}
\end{figure}

\renewcommand{\arraystretch}{1.2} 
\begin{table}
\centering
\resizebox{\textwidth}{!}{%
\begin{tabular}{c|c|c|c|c|c|c}
\toprule

Mocks label & clean & dv-LRG & dv-QSO-G & dv-QSO-L & dv-ELGcatas & dv-slitless \\
\hline
$\sigmadv$/$\wdv$ ($\kmps$) & - & 85.7 & 300 & 100 & - & 85.7 \\
\hline
$\fc$ & - & - & - & - & 1\% & 5\% \\
\bottomrule
\end{tabular}
}
\caption{Summary of contaminations and mock catalogs used in our study. Mocks include uncontaminated (clean) and those affected by redshift errors. We consider redshift uncertainties for LRG-like Gaussian (dv-LRG), QSO-like Gaussian (dv-QSO-G) and QSO-like Lorentzian (dv-QSO-L) profiles. Redshift catastrophic failures are based on analyses of DESI DR1 ELG repeat observations~\cite{yu_elg_2025}, with $\fc$ being the contamination rate. We consider $\fc = 1\%$ with ELG-like catastrophic pattern (dv-ELGcatas) in DESI survey. For slitless-like errors (dv-slitless), we assume combined errors with LRG-like uncertainty and $\fc = 5\%$ ELG-like catastrophics to mimic the expected observing conditions in slitless survey like \textit{Euclid}. The corresponding redshift error distributions applied in contaminated mocks are shown in \cref{fig:hist_model_redshift_errors}.
}
\label{tab:mock catalogs}
\end{table}

We use the \QUIJOTE $N$-body simulation suite to construct our mock catalogs \cite{villaescusa-navarro_quijote_2020}. The \QUIJOTE fiducial simulations assume a flat $\LCDM$ universe, with cosmology consistent with the \textit{Planck} 2018 result \cite{planck_collaboration_planck_2020}: $\h = 0.6711, \, \Om = 0.3175, \, \Ob = 0.049, \, \sigma_8 = 0.834, \, n_s = 0.9624, \, \sumnu=0.0 \, {\rm{eV}}, \, w = -1$. To study the impact of redshift errors on neutrino mass measurement, we also include the ‘\textit{Mnu\_p}’ catalog with $\sumnu = 0.1 \, \mathrm{eV}$. In \QUIJOTE simulation, each realization contains \( 512^3 \) CDM particles with mass $m_p \approx 6.5 \times 10^{11} \, M_\odot$, evolved within a comoving box of volume $1 \, \hgpc$. To match the redshift range of interest, we use the snapshots at redshift $z=1.0$. A mass selection with \( 13.1 < \log_{10}(M_h/(h^{-1}M_\odot)) < 13.5 \) is applied to ensure that the truncated halos represent the dominant structures contributing to the galaxy clustering. We use 500 realizations from the \QUIJOTE\ suite to construct mock catalogs contaminated by different types of redshift errors.

The halo catalogs include information such as the Cartesian positions $X, Y, Z$, the corresponding peculiar velocities \( \bm{v}_{\rm{pec}} \), and the halo masses, \( M_h \). Assuming the line-of-sight is along the $x$-direction, we use the $x$-component of the peculiar velocity, $\vpx$, to transform halo positions from real space to redshift space at redshift $z$: $X_{\rm{rsd}} = X_{\rm{real}} +  \vpx \, (1+z)/H(z)$. The clean mock catalogs are built using this peculiar velocity without modification. To build contaminated mocks with redshift errors, we add an additional velocity shift $\dve$ to $\vpx$. This $\dve$ simulates the line-of-sight velocity dispersion introduced by redshift errors.


To generate the $\dve$ distribution, we use different profiles to simulate redshift errors for different surveys. For redshift uncertainty, we use the magnitudes reported from eBOSS and DESI Early Data Release (EDR) \cite{zarrouk_clustering_2018, yu_desi_2023}. Specifically, we apply Gaussian profiles in \cref{eq:gaussian uncertainty} with $\sigma_{\dv} = 85.7\, \kmps$ to simulate LRG-like smearing and $\sigma_{\dv} = 300 \, \kmps$ for QSO-like smearing. We also consider a Lorentzian profile in \cref{eq:lorentzian uncertainty} to mimic the extended tails observed in QSO redshift uncertainty. The Lorentzian profile is parametrized with $\mathrm{w}_{\dv} = 100 \, \kmps$ and truncated by $\dvs < 3000 \, \kmps$. Using these values, we have almost the same RMS of $\dvs$ for the QSO Gaussian and Lorentzian smearing. The redshift smearing effects for ELG are small ($\sigma_{\dv} \sim 10\, \kmps$; \cite{yu_elg_2025}) and we neglect them in this study. A summary of the configurations for the contaminated mocks is in \cref{tab:mock catalogs}, with the comparison of distributions shown in the left panel of \cref{fig:hist_model_redshift_errors}. In all cases, we assume the redshift errors are symmetric to zero (i.e., $\mu=0$ and $p=0$).

For redshift catastrophics, we adopt a model based on DESI DR1 ELG repeat observations \cite{yu_elg_2025}. The velocity shifts caused by catastrophics follow a symmetric distribution centered at zero, with each half modeled by the profile $\mathcal{LN}(\Delta \hat{v}_c \, | \, 0.64, 0.25^2, 6.62)$ defined in \cref{eq:catastrophics profile}. We set $\fc=1\%$ to represent ELG-like catastrophics, corresponding to the contamination rate in eBOSS and DESI galaxies. As discussed in \cref{sec:data slitless survey}, to mimic redshift errors expected in slitless surveys, we consider a scenario combining LRG-like Gaussian smearing ($\sigmadv = 85.7\, \kmps$) with $\fc=5\%$ ELG-like catastrophics. The contaminated mocks are summarized in \cref{tab:mock catalogs} and the distributions of ELG-like catastrophics and sliltless-like errors are shown in the right panel of \cref{fig:hist_model_redshift_errors}.

\section{Methods}\label{sec:methods}
\subsection{Power spectrum estimator}\label{sec:method power spectrum estimator}

We estimate the power spectrum of mock catalogs using the weighted galaxy density field with a convenient normalization \cite{feldman_power_1994}:
\begin{equation}
F(\textbf{r}) = \frac{w(\textbf{r}) \left[ n_g(\textbf{r}) - \alpha n_s(\textbf{r}) \right]}{A_g^{1/2}}
\end{equation}
where $w(\textbf{r})$ is the weight function. $n_g$ and $n_s$ are the observed number densities of galaxy and random catalogs, respectively. The factor $\alpha = N_g/N_s$ normalizes the contribution from the random, and the normalization factor $A$ is defined with the expected mean space density of galaxies $\bar{n}(\textbf{r})$ as $A_g = \int d\textbf{r} \left[ \bar{n}(\textbf{r}) w(\textbf{r})\right]^2$. The estimator of power spectrum multipoles $P_\ell$ is given by \cite{feldman_power_1994, hand_optimal_2017}: 
\begin{equation}
P_{\ell}(k) = (2\ell + 1)
\int \frac{d\Omega_k}{4\pi}\left[ \int d\textbf{r}_1 \int d\textbf{r}_2 F(\textbf{r}_1) F(\textbf{r}_2) \times e^{i\ \textbf{k} \cdot (\textbf{r}_1 - \textbf{r}_2)} \mathcal{L}_{\ell}(\hat{\textbf{k}} \cdot \hat{\textbf{r}}_h) - \rm{SN}_{\ell} \right]
\end{equation}
Here, $\mathcal{L}_{\ell}$ is the Legendre polynomial and $\textbf{r}_h = (\textbf{r}_1 + \textbf{r}_2) / 2$. The shot noise term $\rm{SN_{\ell}}$ is expressed as ${\rm{SN}}_{\ell}(\textbf{k}) = (1 + \alpha) \int d\textbf{r} \, \bar{n}(\textbf{r}) w^2(\textbf{r}) \mathcal{L}_{\ell}(\hat{\textbf{k}} \cdot \hat{\textbf{r}})$, which only contributes to monopole $(\ell=0)$ and is negligible relative to $P_{\ell}$ for $\ell>0$. 
In practice, we use the \texttt{pypower}\footnote{\url{https://github.com/cosmodesi/pypower}} \cite{hand_optimal_2017} to compute the power spectrum multipoles from our mock catalogs.

\subsection{Galaxy clustering models}

\subsubsection{Effective-field theory of LSS}\label{sec:method EFT of LSS}
Effective field theory ($\EFT$) has become a powerful tool to model the LSS of the Universe. By studying small-scale nonlinear physics, the $\EFT$ of LSS extends the standard perturbation theory to scales beyond the breakdown of the fluid and single-stream approximations~\cite{baumann_cosmological_2012, carrasco_effective_2012, vlah_lagrangian_2015}. With loop corrections and improved modeling of biased tracers with redshift-space distortions, $\EFT$ enables reliable predictions of the galaxy power spectrum multipoles up to $k \approx 0.25 \,\hmpc$ in redshift space \cite{mcdonald_clustering_2009, senatore_redshift_2014, senatore_bias_2015, chen_redshift-space_2021, aviles_redshift_2021}. In general, the galaxy redshift-space power spectrum in $\EFT$ framework is expressed as \cite{desi_collaboration_desi_2025-1}:
\begin{equation}\label{eq:EFT formula}
P^{\rm{\EFT}}_{g, \, s}(k,\mu) =  P_{g, \, s}^{\rm{PT}}(k,\mu) + (\alpha_0 + \alpha_2\mu^2 + \alpha_4\mu^4+...)k^2P_{g, \, s}^{\rm{L}}(k)  + \left( \rm{SN}_0 +\rm{SN}_2 k^2\mu^2 +... \right),
\end{equation}
Here, $\mu$ is the cosine of the angle between the wavenumber vector $\textbf{k}$ and light-of-sight. $P_{g, \, s}^{\rm{PT}}$ is the galaxy redshift-space power spectrum based on perturbation theory (Standard, Euclidian or Lagrangian...). It is derived from the linear matter power spectrum, incorporating loop corrections and galaxy bias parameters, such as $b_1$, $b_2$ and $b_s$. The second terms are the general $\EFT$ counterterms parameterized by $\alpha_i$, which models the non-perturbative physics and the nonlinear mapping between real and redshift space. The last term corresponds to the stochastic contributions characterized by the shot noise parameters $\rm{SN}_i$. The $\EFT$ of LSS approach has been successfully applied in recent analyses of BOSS, eBOSS and DESI, leading to direct constraints on cosmological parameters using the full-shape of power spectrum~\cite{damico_cosmological_2020, desi_collaboration_desi_2025-3}. Several $\EFT$ codes are publicly available~\cite{chen_consistent_2020, damico_limits_2021, noriega_fast_2022}, and have demonstrated consistent performance in the full-shape analysis~\cite{maus_comparison_2024}. In our work, we use the \FOLPS\footnote{\url{https://github.com/henoriega/FOLPS-nu}} \cite{noriega_fast_2022} code to calculate the $\EFT$ power spectrum. 


\subsubsection{Full-Shape fitting approaches}\label{sec:methods fitting approach}

We adopt two fitting strategies to perform full-shape cosmological inference: ShapeFit compression \cite{brieden_shapefit_2021} and Full-Modeling. These approaches have been extensively tested and validated in analyses of the BOSS and DESI surveys \cite{maus_comparison_2023, noriega_comparing_2024, maus_analysis_2024, lai_comparison_2024, desi_collaboration_desi_2025-1}. 

Compression analysis focuses on fitting a set of compressed physical parameters that describe the shape and the anisotropies of the galaxy power spectrum. These include the scaling parameters $\apar$, $\aper$ and the growth rate $f\sigma_8$. The scaling parameters are often reparameterized into the isotropic dilation parameter $\aiso \equiv \apar^{1/3} \aper^{2/3}$ and the Alcock-Paczynski (AP) parameter $\aap \equiv \apar \aper^{-1}$. ShapeFit further introduces two shape-related parameters, $m$ and $n$, to capture information from the early-time transfer function. In this work, we vary only one shape parameter $m$, while keeping $n$ fixed~\cite{brieden_shapefit_2021, desi_collaboration_desi_2025-1}. We report ShapeFit results relative to a reference cosmology, specifically using $dm \equiv m-1$ and $df \equiv f/f^{\rm{fid}}$. 

Another fitting approach, known as Full-Modeling analysis, avoids the compression steps and directly fits cosmological models to the power spectrum signals. In this way, the parameter dependence of the linear power spectrum along and geometrical information are not kept separate. This allows for a direct comparison between theoretical predictions and observational data and thus gives direct constraints on parameters. In our baseline fitting procedure, we consider five cosmological parameters: $h$, $\omega_{\rm cdm}$, $\omega_b$, $\ln(10^{10}A_s)$, $n_s$. These parameters are directly used in theoretical models, which facilitates the calculation and a direct comparison of cosmological biases introduced by redshift errors.

In the context of redshift error studies, ShapeFit provides a direct view of how redshift errors distort the shape of observed power spectrum, as reflected by their impact on compressed physical parameters. This makes the interpretation of redshift errors effect straightforward. The Full-Modeling approach, on the other hand, gives cosmological constraints, allowing us to quantify the biases introduced by redshift errors in final cosmological constraints.

\subsection{Clustering effect of redshift errors}

\subsubsection{Redshift uncertainty damping}\label{sec:method redshift uncertainty damping}

\begin{figure}
    \centering
    \includegraphics[width=\linewidth]{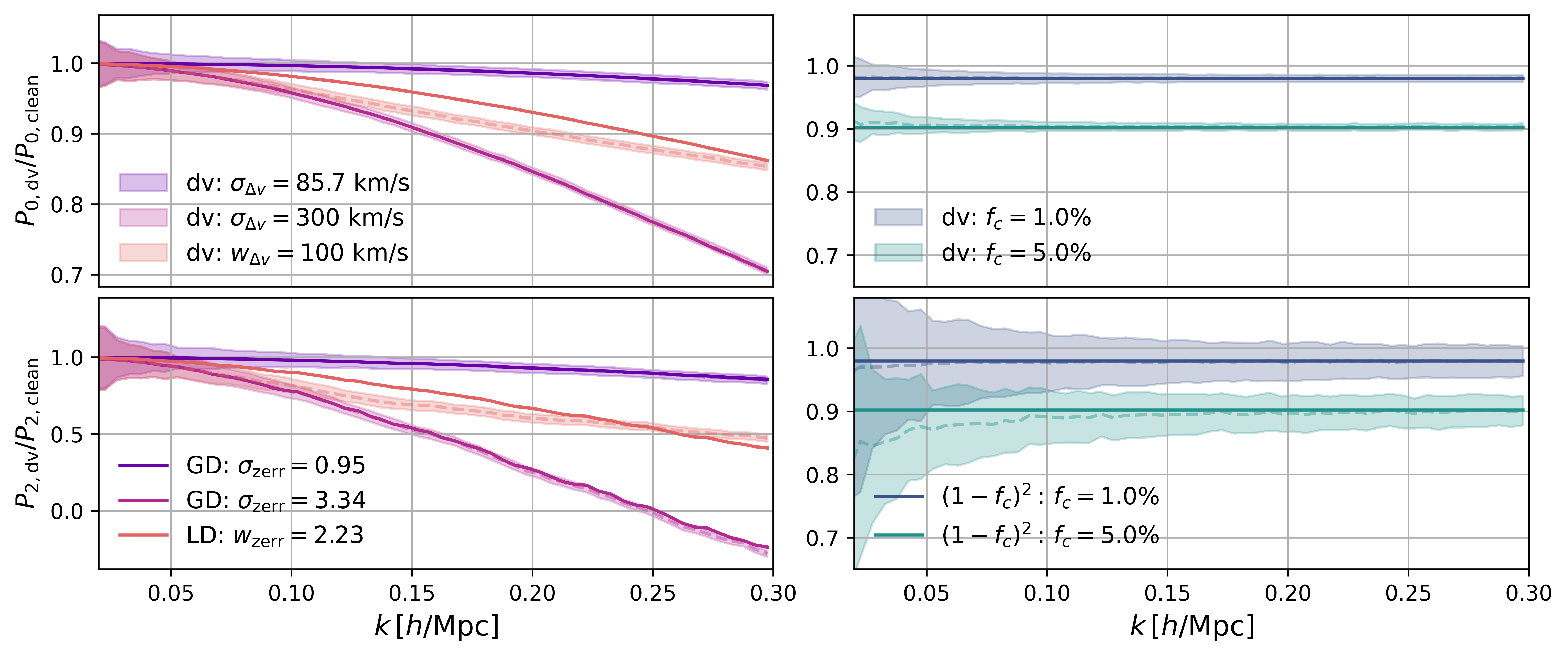}
    \caption{Power spectrum monopole and quadrupole ratios from the mean of 500 mocks (dashed line) compare to theoretical model (solid lines). The shaded regions correspond to the errors from the standard deviation of the 500 mocks divided by 5. \textit{Left}: Comparison between the ratios from mocks contaminated by Gaussian or Lorentzian smearing (characterized by $\sigmadv$ and $\wdv$ respectively) relative to clean mocks with the Gaussian Damping (GD) or Lorentzian Damping (LD) factor. \textit{Right}: Comparison between $\fc=1\%$ and $5\%$ contamination mock ratios with the $(1-\fc)^2$ scaling.}
    \label{fig:plot_pk_ratio}
\end{figure}

As pointed out in the eBOSS QSO analyses \cite{smith_completed_2020, hou_completed_2020}, an approach to model the impact of redshift uncertainty in power spectrum is to introduce an additional damping term $D(k, \mu)$:
\begin{equation}\label{eq:Gaussian damping}
    P_{g, \, s}^{\rm{smear}}(k, \mu) = P_{g, \, s}(k, \mu) \, D(k, \mu)
\end{equation}
\begin{equation}
    D(k, \mu) = 
\begin{cases} 
    \exp \! \left[ -(k \mu \sigmaz)^2 \right], & \text{Gaussian damping}, \\
    1/\left[1+(k \mu \wz)^2\right], & \text{Lorentzian damping}.
\end{cases}
\end{equation}
where the redshift error parameters $\sigmaz$ or $\wz$ quantify the scale of redshift uncertainty. For the Gaussian smearing case, $\sigmaz$ is related to the smearing value with $\sigmaz = \sigmadv * {(1+z)}/{H(z)}$, and, as shown in the left panels of \cref{fig:plot_pk_ratio}, this smearing effects are well captured by the Gaussian damping factor up to $k\sim0.3 \, \hmpc$. In the Lorentzian smearing case, the parametrization implies an additional factor of two, i.e. $\wz = 2\, \wdv * {(1+z)}/{H(z)}$. With Lorentzian damping, we see deviations compared to our mock analyses (orange lines in \cref{fig:plot_pk_ratio}). This is possibly due to the velocity cut we applied when constructing the Lorentzian smearing profile. However, as we will demonstrate in the following, these deviations can be effectively absorbed within the more flexible $\EFT$ framework. 

To compare the damping factor with $\EFT$ model, we can apply the Taylor expansion on the damping terms. For both Gaussian and Lorentzian smearing, we have:
\begin{equation}
    D(k, \mu) \, \rightarrow  1 + a_2 (k\mu)^2 + a_4(k\mu)^4 + \ldots
\end{equation}

We notice that the second and third terms scale as $P(k)(k\mu)^2$ and $P(k)(k\mu)^4$, respectively. These terms have similar structures to those of $\EFT$ counterterms in \cref{eq:EFT formula}. In the $\EFT$ of LSS, the general form of such counterterms can be derived from a moment expansion of the redshift-space power spectrum~\cite{senatore_redshift_2014, aviles_redshift_2021, damico_taming_2024}, as well as from comparisons with phenomenological models~\cite{ivanov_cosmological_2020}. Consistently, \cite{simon_cosmological_2023} has shown that the leading corrections coming from redshift uncertainty are degenerate with $\EFT$ counterterms, particularly those parameterized by $\alpha_2$ and $\alpha_4$. In \cref{sec:result all fits}, we demonstrate equivalently that the redshift uncertainty effect can be absorbed by the $\EFT$ nuisance parameters. Therefore, when we do the cosmological fitting, we use the $\EFT$-based clustering model without including the additional damping factor.


\subsubsection{Redshift catastrophics correction}\label{sec:method redshift catastrophics correction}

Redshift catastrophics have been addressed in theoretical models for redshift surveys like WiggleZ \cite{blake_wigglez_2010} and more recently \textit{Euclid} \cite{euclid_collaboration_euclid_2025-5}. Assuming that the redshift catastrophic contamination rate is $\fc$, an approximate correction is to multiply the $\EFT$ galaxy power spectrum by a factor of $(1-\fc)^2$: 
\begin{equation}\label{eq:catastrophics simple model}
P^{\EFT+\fc}_{g, \, s}(k, \mu) \approx (1 - \fc)^2 P^{\EFT}_{g, \, s}(k, \mu)
\end{equation}
The validity of this approximation depends both on $\fc$ and the displacement caused by catastrophics. In \cref{fig:plot_pk_ratio}, based on our mock analyses with contamination $\fc=1\%$ and $5\%$, we find that the relative difference between the correction factor, $R_{\rm{corr}} \equiv (1 - \fc)^2$, and the ratio from the mock statistics, $R_{\rm{mock}} \equiv P_{\text{dv}, \, \fc}/P_{\rm{clean}}$, satisfies:
\begin{equation}
\Bigg\lvert\frac{R_{\rm{corr}}(k)-R_{\rm{mock}}(k)}{R_{\rm{mock}}(k)} \Bigg\rvert \lesssim 6\%, \quad \rm{For} \ 0.02 \leq k \leq 0.3 \, \hmpc
\end{equation}
This suggests that the approximation is sufficiently accurate for full-shape power spectrum analysis, especially considering that additional nuisance parameters in the $\EFT$ can absorb residual mismatches. In DESI, the $\fc$ is typically below 1\%. Their overall impact is expected to be small \cite{yu_elg_2025} and such correction is not included in the standard $\EFT$ model. However, in slitless surveys such as \textit{Euclid}, where $\fc$ is significantly higher, we will show in~\cref{sec:result fit with fc} that incorporating this correction will be necessary to obtain unbiased cosmological constraints.

\subsection{Cosmological inference and Covariance matrices}\label{sec:method covariance}

Cosmological inference employs Bayesian statistics to infer constraints on physical or cosmological parameters. Assuming the evidence from the data is normalized to one, the posterior probability is expressed as $P(\Theta|\mathcal{D}) \propto \mathcal{L}(\mathcal{D}|\Theta)P(\Theta)$, with $\Theta$ being the set of input parameters, including compressed parameters $\left\{ \aiso,  \; \aap,  \; df,  \; dm  \right\}$ by ShapeFit or cosmological parameters $\left\{ h, \; \omega_{cdm}, \; \omega_{b},  \; \rm{ln}(10^{10}A_s), \; n_s \right\}$ by Full-Modeling. It also contains nuisance parameters considered in the $\EFT$ model: $\left\{ b_1, \; b_2, \; b_s,  \; \alpha_0, \; \alpha_2, \; \alpha_4, \; \rm{SN}_0, \; \rm{SN}_2 \right\}$. Assuming that we have Gaussian-distributed data, the likelihood function can be expressed as $\mathcal{L}(\Theta) \propto e^{-\chi^2(\Theta)/2}$. The $\chi^2$ statistic quantifies the goodness of fitting and is computed based on the difference between the observed data and the model predictions, which reads as:
\begin{equation}
    \chi^2(\Theta) = (\mathcal{D}_{\rm{data}}-\mathcal{D}_{\rm{model}}(\Theta))^T \, \mathcal{C}^{-1} \, (\mathcal{D}_{\rm{data}}-\mathcal{D}_{\rm{model}}(\Theta))
\end{equation}
In our study, the vector $\mathcal{D}$ includes a combination of the power spectrum monopole  $P_0(k)$ and quadrupole $P_2(k)$. The subscript 'data' and 'model' refers to the measurements from observations (here mocks) and theoretical model predictions. For each type of redshift-error, we have constructed 500 corresponding contaminated mocks and use them to build the data vector and covariance matrix. The covariance matrix $\mathcal{C}$ is computed and normalized to the number of the datasets, $N_d$, with the formulation:
\begin{equation}
    \mathcal{C}_{ij} = \frac{1}{N_d - 1} 
    \sum_{n=1}^{N_d} 
    \left( \mathcal{D}_i^{(n)} - \overline{\mathcal{D}}_i \right)
    \left( \mathcal{D}_j^{(n)} - \overline{\mathcal{D}}_j \right),
\end{equation}
The $\mathcal{D}_i^{(n)}$ denotes the $i$-th bin of the data vector from the $n$-th mock realization and $\overline{\mathcal{D}}_i$ is the ensemble mean. In \QUIJOTE suite, the volume of each realization is $V_1 = 1 \, \hgpc$. To model the reduced cosmic variance expected from galaxy surveys, we use the rescaled covariance as: $\mathcal{C'} = ({V_1}/{V_n}) \, \mathcal{C}$, where $V_n$ is the target survey volume. In our analyses, we rescale the covariance to an effective volume of $V_{25} = 25 \,\hgpc$ to match the expected survey volumes of DESI and \textit{Euclid}.

\section{Results}\label{sec:results}
\subsection{Impact of redshift errors on clustering statistics}\label{sec:result clustering impact}

\begin{figure}
    \centering
    \includegraphics[width=\linewidth]{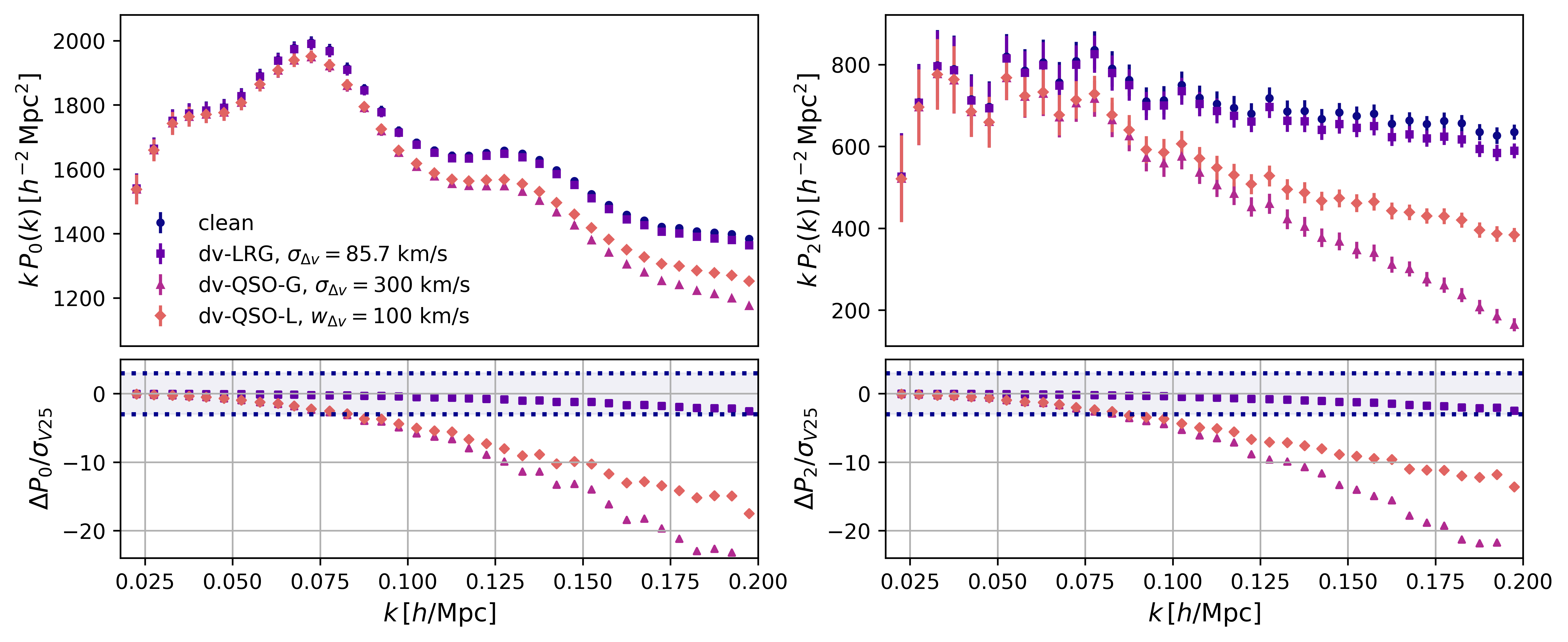}
    \caption{Comparison of power spectrum monopole (left) and quadrupole (right) between clean mocks and mocks contaminated by redshift uncertainties. The error bars, $\sigma$, are rescaled to match an effective survey volume of $V_{25}$. The lower panels show their residuals in units of clean mocks $\sigma$, with the dotted line indicating the $3\sigma$ threshold. In QSO-like Gaussian or Lorentzian smearing cases, we see significant deviations ($>3\sigma$) for $k \gtrsim 0.09 \, \hmpc$.}
    \label{fig:plot_clustering_pk_ru}
\end{figure}

\begin{figure}
    \centering
    \includegraphics[width=\linewidth]{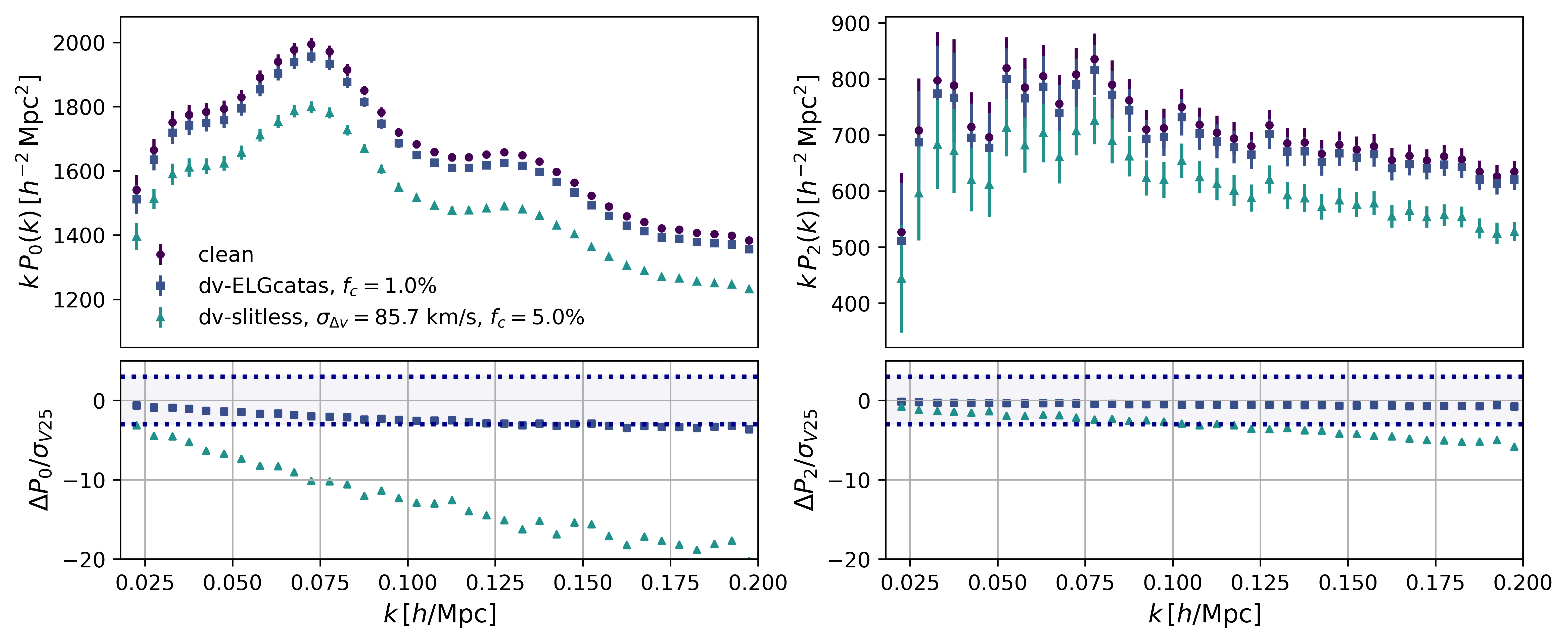}
    \caption{Similar to \cref{fig:plot_clustering_pk_ru} comparing clean mocks and mocks contaminated by ELG-like catastrophics and slitless-like errors. Redshift catastrophics introduce an overall suppression on power spectrum amplitude, leading $>3\sigma$ difference for sliltess-like errors.}
    \label{fig:plot_clustering_pk_rc}
\end{figure}

The impact of redshift errors on the clustering statistics is presented in \cref{fig:plot_clustering_pk_ru} and \cref{fig:plot_clustering_pk_rc}. For redshift uncertainty, it introduces a smearing effect that systematically suppresses the power spectrum amplitude, with the suppression becoming stronger at smaller scales (i.e., larger $k$). The lower panels of \cref{fig:plot_clustering_pk_ru} show the residual plot in units of the clean mocks error bar $\sigma$. For an effective survey volume of $V_{25}$, the LRG-like smearing leads to minimal impacts on the clustering, with power spectrum difference $\Delta P < 3\sigma$. For QSO-like redshift uncertainty, both Lorentzian and Gaussian smearing lead to $\Delta P>3\sigma$ for $k \gtrsim 0.09 \, \hmpc$ in the monopole and quadrupole. Notably, the QSO Lorentzian smearing, with the same RMS value as the Gaussian case, shows a comparatively flatter suppression at large $k$.

Redshift catastrophics affect only a small fraction of objects but cause significant shifts on their positions along the line-of-sight. \Cref{sec:method redshift catastrophics correction} shows that this distortion leads to an overall reduction on the power spectrum, scaling approximately as $(1 - \fc)^2$. As illustrated in \cref{fig:plot_clustering_pk_rc}, the ELG-like catastrophics with $\fc = 1\%$ leads to a 2\% reduction of the clustering amplitude, with maximum deviations around $3 \sigma$. However, slitless-like errors (i.e., $\sigma_v=85.7 \, \kmps$ and $\fc = 5\%$) introduce an overall suppression of the power spectrum amplitude by 10\%-12\%, resulting in a deviation $>3\sigma$ for $k \gtrsim 0.02 \, \hmpc$ in the monopole and $k\gtrsim0.12 \, \hmpc$ in the quadrupole. Such $3\sigma$ deviations, if uncorrected, could bias cosmological inferences. In the following subsections, we will describe their impact when running the inference in different cosmological cases: $\LCDM$, $\wCDM$ and massive neutrino.


\subsection{Baseline Fitting}

Our baseline full-shape analysis fits the power spectrum monopole and quadrupole over the range $ 0.02  \leq k\leq 0.2\, \hmpc$, using the standard $\EFT$ model without including the damping term from redshift uncertainty or the correction factor from redshift catastrophics. We adopt two fitting strategies, ShapeFit and Full-Modeling, to extract cosmological information. The results are presented within the standard flat-$\LCDM$ model assuming a BBN prior on baryon abundance $\Ob$ and wide Gaussian prior on spectral index $\ns$. The cosmological inference is performed using the \desilike\footnote{\url{https://github.com/cosmodesi/desilike}} package and parameter priors are set to the same as the those in DESI DR1 full-shape analysis \cite{desi_collaboration_desi_2025-1}. The parameter space is sampled with the code \emcee~\cite{foreman-mackey_emcee_2013}. All plots and confidence intervals are generated using the \getdist\, package~\cite{lewis_getdist_2025}. Beside the baseline full-shape analysis, we also evaluate the impact of redshift errors on BAO fitting. These results are discussed in \cref{Apd:BAO fitting}.

\subsubsection{Cosmological influence}\label{sec:result all fits}

\begin{figure}
    \centering
    \includegraphics[width=\linewidth]{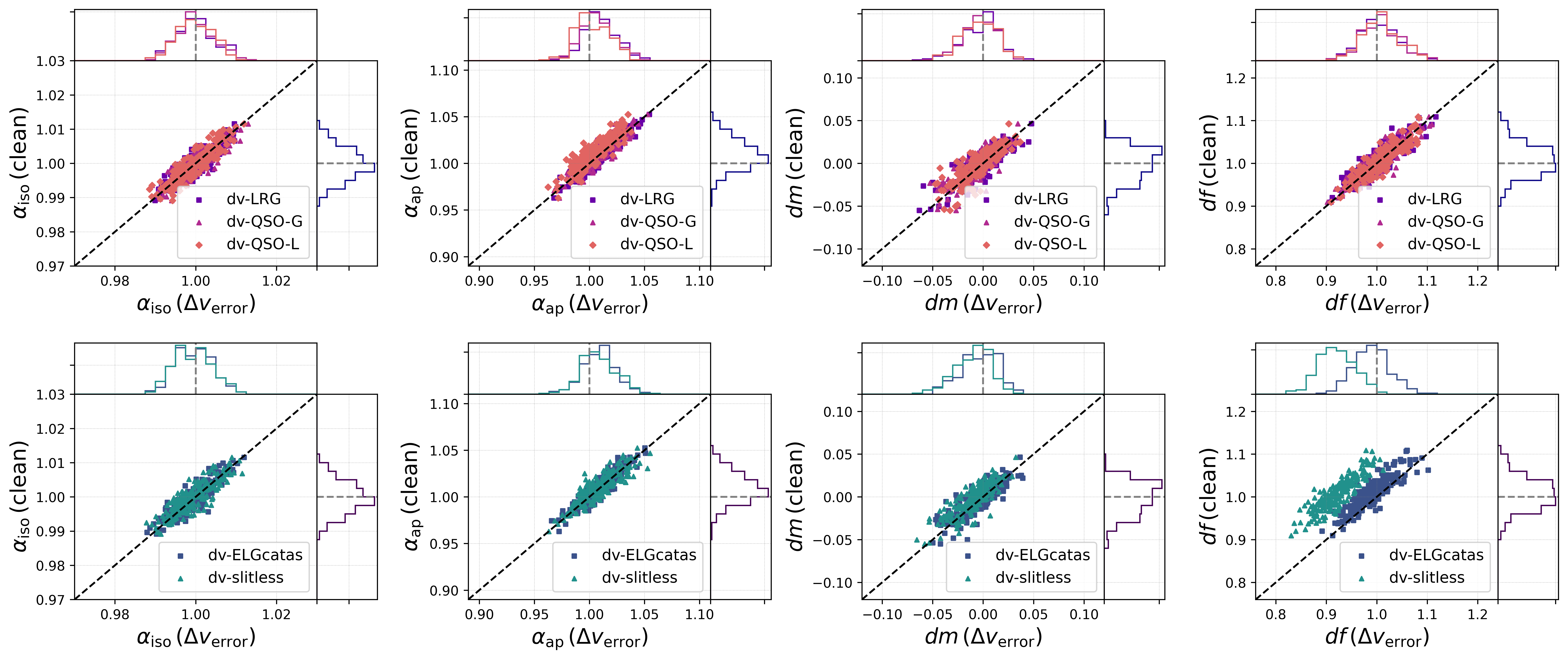}
    \caption{Distributions of the best-fit values for ShapeFit compressed parameters. Scatter plots compare results from clean and contaminated mocks in \cref{tab:mock catalogs}. Each point represents one fit to the mean of 25 random mocks, with the covariance rescaled to volume $V_{25}$. The corresponding marginalized histograms are shown in the right and top subpanels. The parameter estimates remain consistent across most cases, except for $df$, which shows a notable deviation in mocks contaminated by slitless-like errors.}
    \label{fig:scatter_SF_bestfit}
\end{figure}

\begin{figure}
    \centering
    \includegraphics[width=0.9\linewidth]{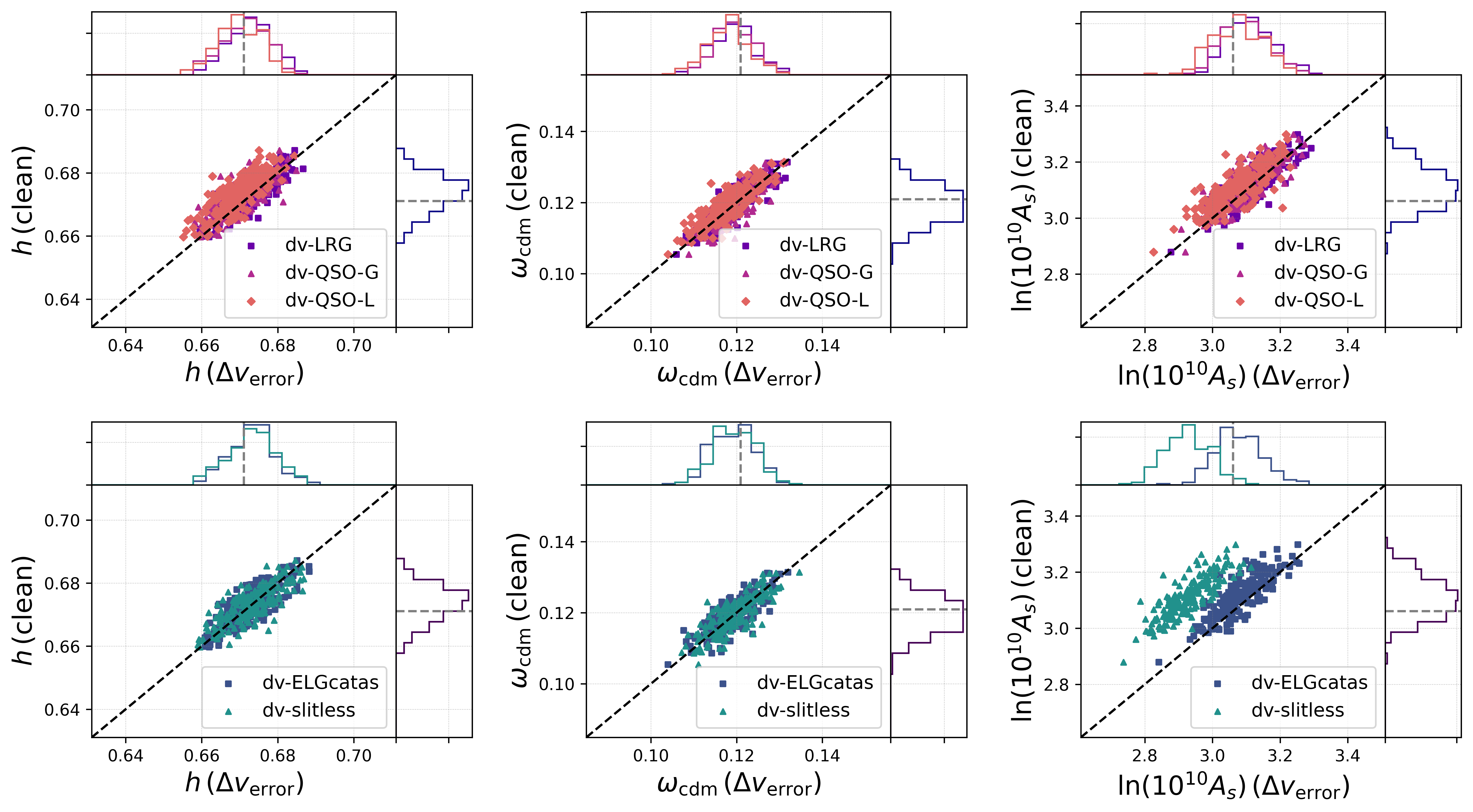}
    \caption{Same plots as \cref{fig:scatter_SF_bestfit} for Full-Modeling cosmological parameters. As before, results from clean and contaminated mocks in \cref{tab:mock catalogs} are compared. Only $\logA$ shows a notable deviation in mocks contaminated by slitless-like errors.}
    \label{fig:scatter_FM_bestfit}
\end{figure}

\begin{figure}
    \centering
    \includegraphics[width=1.0\linewidth]{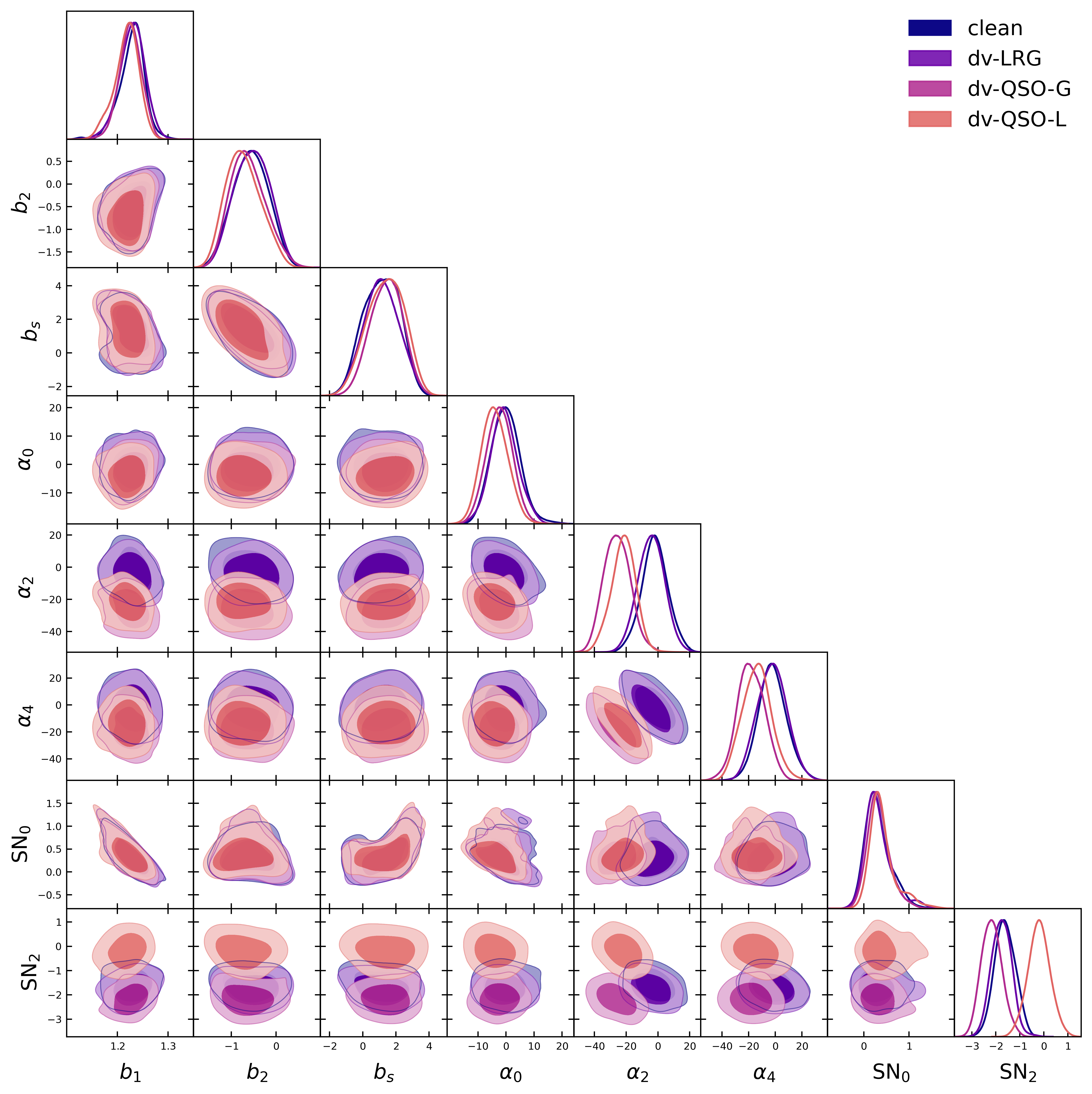}
    \caption{Posterior distribution for nuisance parameters in our baseline \EFT~fitting. Constraints are compared from clean mocks (dark blue) and those contaminated by redshift uncertainties: LRG-like Gaussian (purple), QSO-like Gaussian (pink) and QSO-like Lorentzian (red) smearings. Contours show the 68\% and 95\% confidence level. Results are derived from fits to the mean power spectrum monopole and quadrupole of 25 mocks using covariance rescaled to $V_{25}$. The marginalized posteriors of $\alpha_2$ and $\alpha_4$ show negative shifts, reflecting the smearing effect of redshift uncertainty, while the positive shift of $\mathrm{SN}_2$ shows the difference brought by the Lorentzian smearing.}
    \label{fig:corner_nuissance_ru}
\end{figure}

In this section, we assess the impact of redshift errors on cosmological inference. These results are based on 200 independent fits to the \QUIJOTE mocks catalogs in \cref{tab:mock catalogs}. For each fit, we use the mean power spectrum derived from 25 randomly selected mocks (drawn from 500), with a covariance matrix rescaled to the volume $V_{25}$. Comparisons of the best-fit parameters between analyses from clean and contaminated mocks are presented in \cref{fig:scatter_SF_bestfit} for ShapeFit and in \cref{fig:scatter_FM_bestfit} for Full-Modeling. The corresponding $\chi^2$ comparison, provided in \cref{Apd:Supplementary figures}, confirms that all our fits achieved comparable goodness-of-fit.

In \cref{fig:scatter_SF_bestfit} and \cref{fig:scatter_FM_bestfit}, the upper panels illustrate the impact of redshift uncertainty on cosmological inference. The fitting results from both ShapeFit and Full-Modelling indicate that no significant bias (<5\%) in cosmological parameter estimates. As discussed in \cref{sec:method redshift uncertainty damping}, the suppression of the power spectrum caused by redshift uncertainty is expected to be absorbed by $\EFT$ counterterms. \cref{fig:corner_nuissance_ru} plots the posterior distributions of all nuisance parameters. These posteriors reveal negative shifts in $\alpha_2$ and $\alpha_4$, consistent with the expected absorption of the damping effect. Another parameter that shows variation is $\mathrm{SN}_2$, which represents the second-order stochastic term and scales as $(k\mu)^2$. Its positive shift in the Lorentzian smearing case could be led by the flatter suppression pattern of the Lorentzian profile at large $k$.

The lower panels of \cref{fig:scatter_SF_bestfit} and \cref{fig:scatter_FM_bestfit} assess the impact of redshift catastrophics and hypothetical slitless-like errors on cosmological inference. The ELG-like catastrophics with $\fc=1\%$ show negligible bias in parameter estimates (with the largest discrepancy $<6\%$), consistent with the previous findings in \cite{yu_elg_2025}. In contrast, slitless-like errors with $\fc=5\%$ show noticeable biases, particularly in estimating the fractional growth rate parameter $df$ in ShapeFit and the log primordial amplitude parameter $\logA$ in Full-Modeling. Both parameters are systematically underestimated by $6\%{-}16\%$ compared to clean case. We identify that this negative shift arises mainly from the $5\%$ catastrophic contamination in slitless-like errors, which significantly suppresses the power spectrum amplitude (see \cref{fig:plot_clustering_pk_ru}). If not adequately modeled in theory, this effect could bias the estimates of amplitude-related parameters. Nevertheless, in \cref{sec:method redshift catastrophics correction}, an approximate correction by multiplying the power spectrum by a factor of $(1 - \fc)^2$ can mitigate this suppression. In the following section, we incorporate this correction into the cosmological analyses to further investigate the influence of redshift catastrophics. 


\subsubsection{Catastrophics-correction for slitless survey}
\label{sec:result fit with fc}

\begin{figure}
  \centering
  \begin{subfigure}[b]{0.496\textwidth}
    \includegraphics[width=\textwidth]{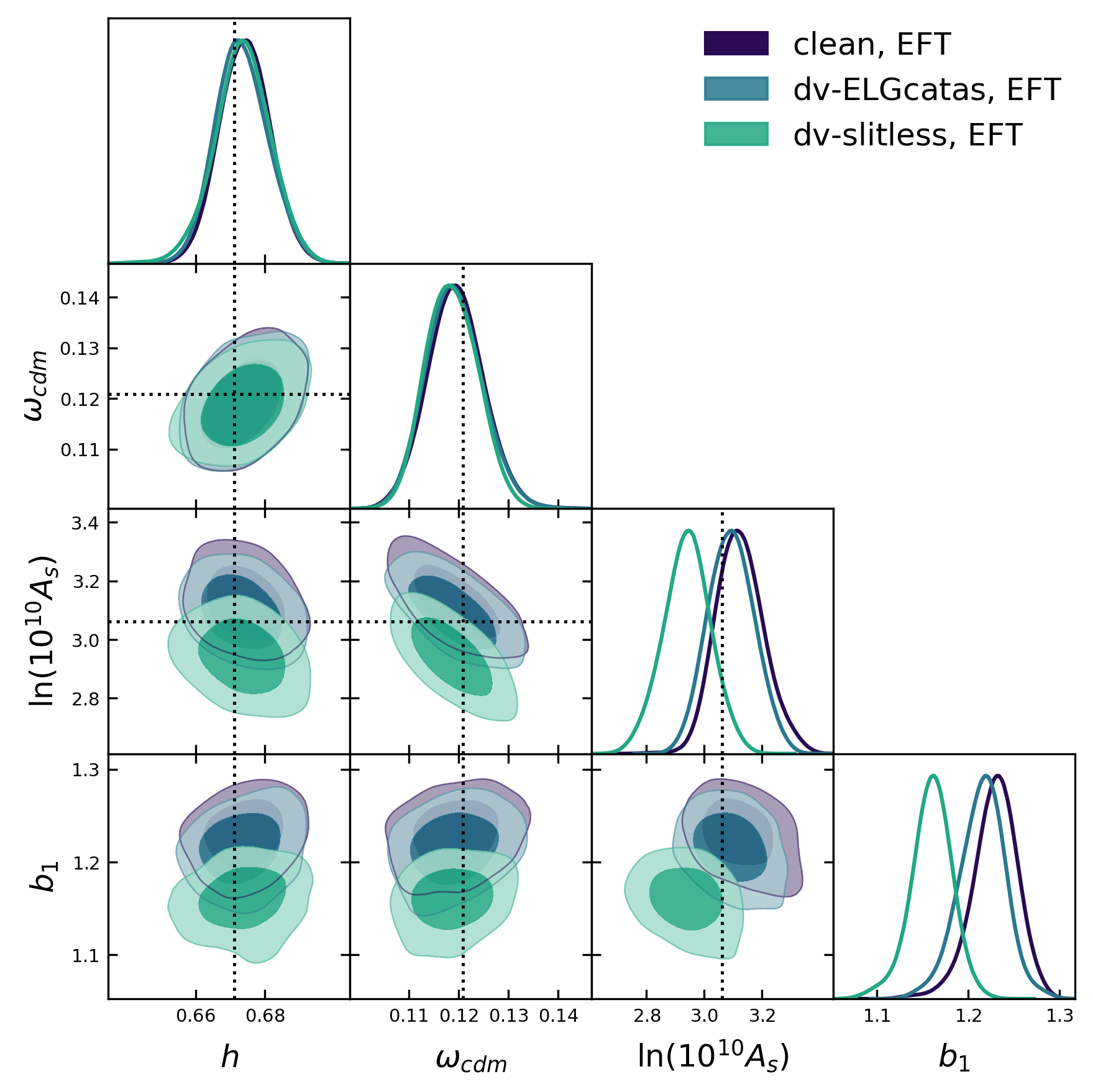}
  \end{subfigure}
  \hfill
  \begin{subfigure}[b]{0.496\textwidth}
    \includegraphics[width=\textwidth]{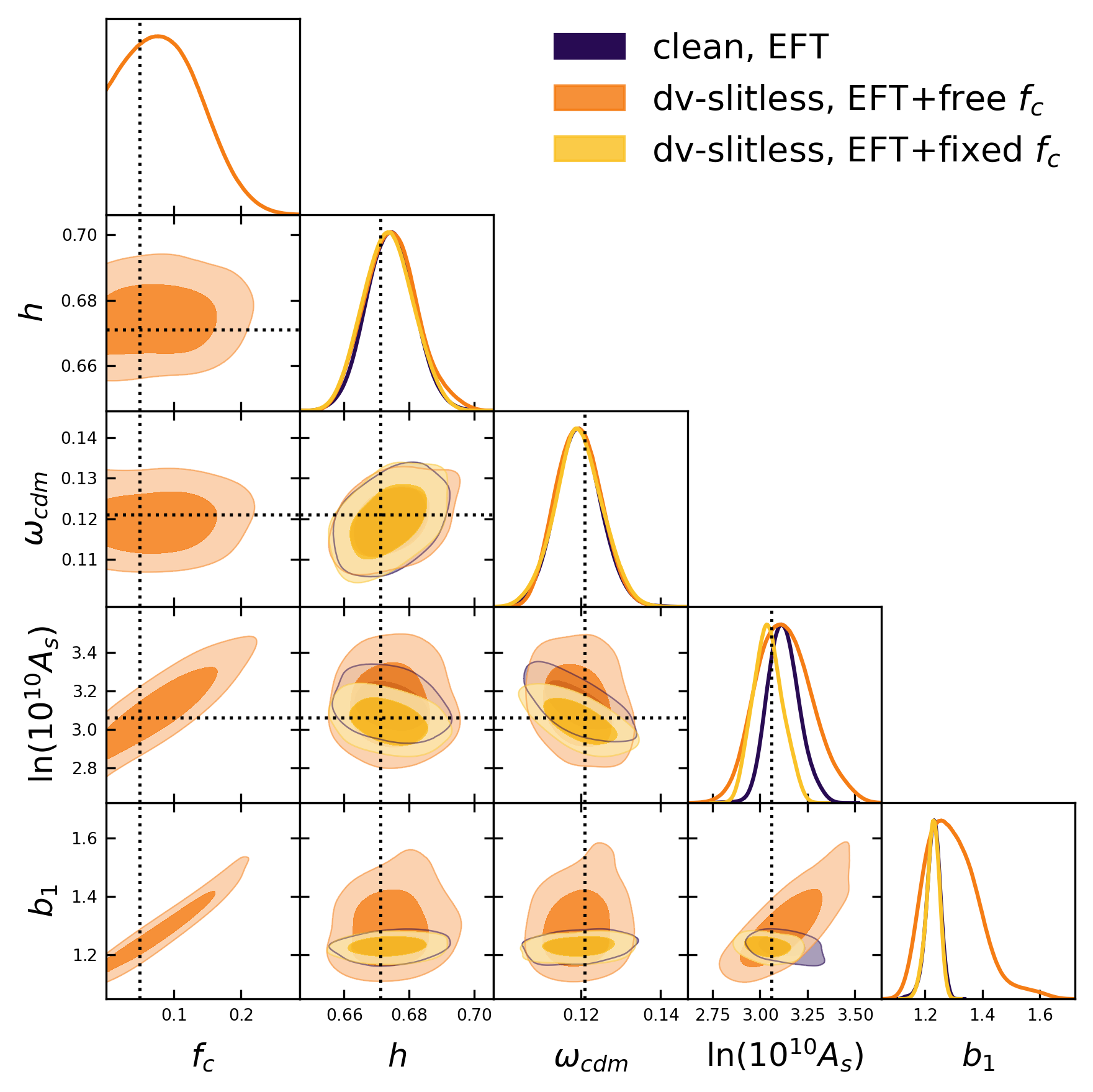}
  \end{subfigure}
  \caption{Posterior distributions for Full-Modeling parameters. \textit{Left panel}: Constraints using the baseline $\EFT$ model to fit clean mocks (dark blue), mocks contaminated by ELG-like catastrophics ($\fc=1\%$; blue), and slitless-like errors ($\sigma_v=85.7\kmps$ with $\fc=5\%$; green). \textit{Right panel}: Comparison between clean results with slitless-like errors using the extended $\EFT$ model when freeing $\fc$ ($\EFTfreefc$; orange) and with $\fc$ fixed to its expected value ($\EFTfixedfc$; yellow). Contours show 68\% and 95\% confidence intervals with black dash lines indicating the true values. Results are derived from fits to the mean power spectrum monopole and quadrupole of 500 mocks using the covariance matrix rescaled to $V_{25}$. The marginalized posteriors of $\logA$ and $b_1$ are shifted by slitless-like errors, but using the corrected model with $\fc$ help restore the estimated values.}
  \label{fig:corner_fc_model}
\end{figure}

\begin{table}
\centering
\setlength\tabcolsep{5.5pt}
\renewcommand\arraystretch{1.2}
\makebox[\textwidth][c]{
\begin{tabular}{c|c|c|c|c|c}
\toprule
Theory & \multicolumn{3}{c|}{\EFT~baseline} & \multicolumn{1}{c|}{$\EFTfreefc$} & \multicolumn{1}{c}{$\EFTfixedfc$} \\
\hline
Mocks & clean & dv-ELGcatas & dv-slitless & dv-slitless & dv-slitless\\
\hline
\hline
$\ln(10^{10} A_{s})$ &  $3.125\pm0.084$  &  $3.095\pm0.081$ &  $2.941\pm0.085$  & $3.13\pm0.14$ &  $3.042\pm0.079$ \\
\bottomrule
\end{tabular}}
\caption{Constraints on $\logA$ for clean and contaminated mocks fitted by different theoretical models. Table shows the mean values with 68\% confidence intervals. In $\EFT$ baseline fit, the mean values from ELG-like catastrophics and sliltless-like errors show deviations of $0.4\sigma$ and $2.2\sigma$, respectively, compared to the clean case. By multiplying the factor $(1-\fc)^2$ to the galaxy power spectrum, the extended model $\fc$, $\EFTfreefc$ recovers an unbiased value of $\logA$ but increases its error bar by 67\% due to parameter degeneracies. Fixing the $f_c$ to its expected value, the $\EFTfixedfc$ model restores the constraining power with a modest bias of $1.0\sigma$.}
\label{tab:1D constrain on logA}
\end{table}

In this test, we present Full-Modeling fitting results based on the mean power spectrum calculated from all 500 mocks, with the covariance matrix rescaled to volume $V_{25}$. By averaging over 500 mocks, we minimize the impact of cosmic variance, so any observed deviations in parameter estimates can be more confidently attributed to redshift errors. The left panel of \cref{fig:corner_fc_model} shows the posterior distributions of parameters using the baseline $\EFT$ fit for three contamination cases: clean, ELG-like catastrophics ($\fc = 1\%$), and slitless-like errors (i.e., $\sigma_v=85.7\kmps$ and $\fc = 5\%$). The corresponding mean values of $\logA$ with 68\% confidence intervals are summarized in \cref{tab:1D constrain on logA}. Comparing the $\logA$ constraint from clean case, we find a small shift of $0.4\sigma$ by ELG-like catastrophics and a significant one of $2.2\sigma$ by slitless-like errors. These results are consistent with our earlier findings in \cref{sec:result all fits}. The linear bias parameter $b_1$ is also known to be related to the overall amplitude of galaxy power spectrum. In \cref{fig:corner_fc_model}, we find a shift of $2.7\sigma$ in its mean value for the slitless-like errors case using baseline \EFT. We note that $b_1$ is highly degenerate with $f_c$ and $\logA$, but the shift in $b_1$ is insufficient to compensate for the bias induced by slitless-like errors on $\logA$.

To incorporate the effect of catastrophics in the clustering model, we extend the baseline $\EFT$ model by adding an additional parameter, catastrophic rate $\fc$, and multiplying the factor $(1-\fc)^2$ to the galaxy power spectrum as described in \cref{eq:catastrophics simple model}. When freeing $\fc$ parameter during sampling with a flat prior $\fc \in [0,1]$, we refer to the $\EFTfreefc$ model. The posterior distributions, including the variation of $\fc$, are shown in the right panel of \cref{fig:corner_fc_model}, with the corresponding constraint on $\logA$ reported in \cref{tab:1D constrain on logA}. The $\EFTfreefc$ model succeed in recovering an unbiased mean value of $\logA$ for slitless-like errors, with a discrepancy below $0.1\sigma$ compared to the clean case. However, this improvement comes at the cost of loss in constraining power: we measure a $\sim67\%$ degradation in the $\logA$ constraint. This loss is due to the strong degeneracy between $\fc$ and $\logA$, as illustrated in the \cref{fig:corner_fc_model} $\logA{-}\fc$ plane. When fixing the parameter $\fc$ to its expected value, referred to the $\EFTfixedfc$ model, we successfully recover the constraining power but find a slightly larger deviation ($\sim1.0\sigma$) in the estimated mean value. This deviation is possibly due to the combined effects of redshift uncertainty in the slitless-like errors and the approximation applied in the correction model.

This test is informative for slitless spectroscopy survey like \textit{Euclid}, which aims to constrain cosmological parameters via galaxy power spectra. As we have shown that this hypothetical redshift errors can bias cosmological inference, an accurate estimation of the catastrophics rate and the implementation of the catastrophics-corrected model, either with $\fc$ fixed or varied under a narrow Gaussian prior, are necessary to get unbiased constraints. Since our slitless-like errors scenario is optimistic and slitless survey may have other additional complex systematics, further investigations on modeling redshift errors are still required.


\subsection{Beyond $\LCDM$ cosmologies}

Recent spectroscopic surveys such as DESI have extended full-shape power spectrum analysis beyond the standard $\LCDM$ framework to test cosmological models with evolving dark energy ($\wCDM$) and massive neutrinos. Assuming a flat Universe and combining the full-shape and BAO analyses with external datasets, the DESI DR2 finds a mild preference for a dynamical dark energy model with $w_0>-1$ and $w_a<0$ ~\cite{desi_collaboration_desi_2025-4} and places a tight constraint on the total neutrino mass, $\sumnu<0.0642 \, \rm{eV} \ (95\% CL)$ \cite{elbers_constraints_2025}. This section investigates how redshift errors affect the measurements of dark energy equation of state $w_0$-$w_a$ and $\sumnu$ through the full-shape analysis. In the following, we use the baseline $\EFT$ fitting setup and present results obtained by fitting the mean power spectrum of 500 mocks, with the covariance matrix rescaled to the effective volume $V_{25}$.

\subsubsection{Evolving dark energy}\label{sec:result wowaCDM}


\begin{figure}
    \centering
    \includegraphics[width=1.0\linewidth]{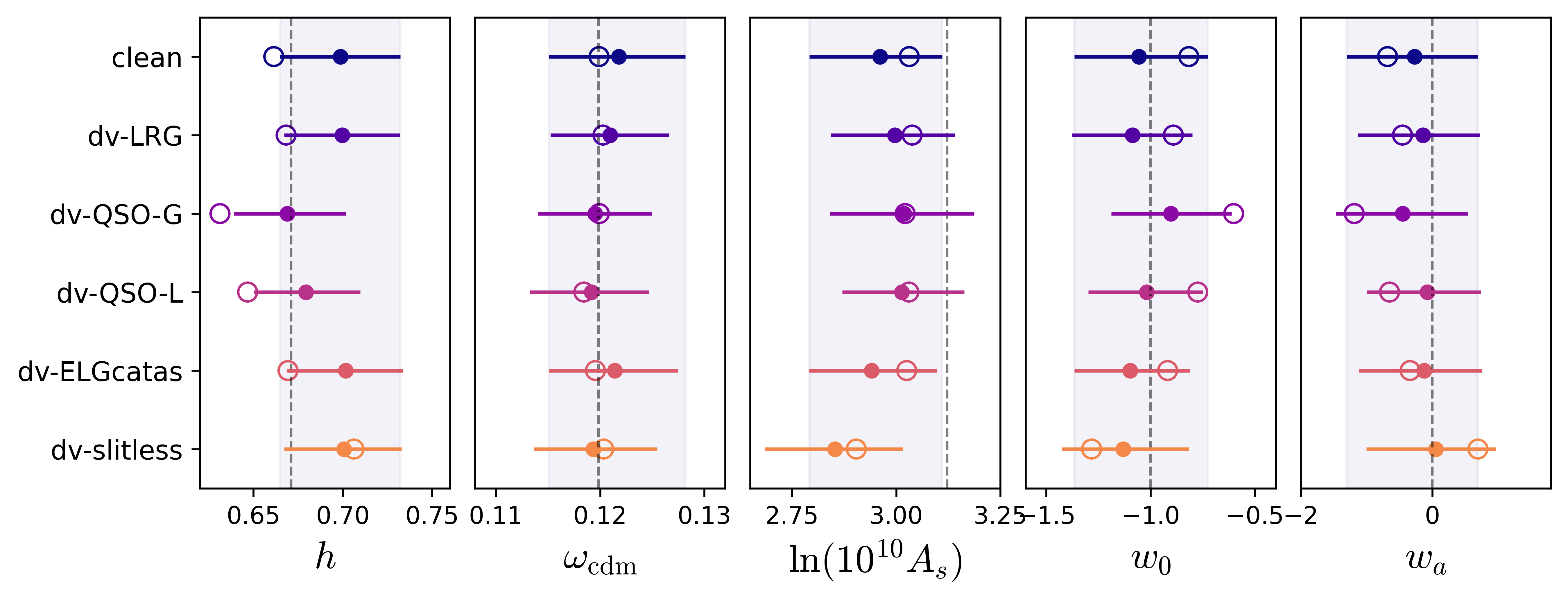}
    \caption{Parameter constraints on $\wCDM$ model for clean and contaminated mocks. The horizontal error bars indicate mean and the 68\% credible levels of the corresponding posterior. The shaded region show the $1\sigma$ constraints from the clean mocks. While redshift errors do not introduce bias in the measurement of dark energy properties $w_0$, $w_a$, they potentially lead to stronger projection effects, which are deviations between the MAP values (open circles) with the mean of the posterior, particularly with QSO-like smearing.}
    \label{fig:1d_constraint_w0wa}
\end{figure}

We assess the influence of redshift errors on the measurement of evolving dark energy using the Chevallier-Polarski-Linder (CPL) parametrization $w(a) = w_0 + w_a (1-a)$ \cite{chevallier_accelerating_2001, linder_exploring_2003}. The cosmological fits are performed using the \QUIJOTE fiducial mocks described in \cref{sec:data contaminated mocks}, which assume a $\LCDM$ model with $w_0=-1$ and $w_a=0$. The resulting parameter constraints are shown as the horizontal lines in \cref{fig:1d_constraint_w0wa}. From the comparisons, we find that redshift errors do not introduce significant bias in the measurement of dark energy properties. However, we know that using full-shape analysis alone to constraint $\wCDM$ model suffer from projection effects \cite{desi_collaboration_desi_2025-1}. To study this aspect, we use $\iminuit$ \cite{james_minuit_1975} to estimate the maximum a \textit{posteriori} (MAP) value, starting from the maximum log-posterior points found in MCMC chains. In \cref{fig:1d_constraint_w0wa}, we find offsets between the mean of the marginalised posteriors (solid circles) and MAP values (open circles) for parameters $h$, $w_0$ and $w_a$, with the offsets being most pronounced for mocks contaminated by QSO-like smearing. Since the redshift errors impact the clustering pattern similarly with the behavior of certain parameters (cosmological or nuisance), their presence could complicate the parameter degeneracies and thus amplify projection effects. Such impact should be investigated in more detail for redshift surveys aiming to constrain dark energy by full-shape analysis.


\subsubsection{Massive neutrinos}\label{sec:result nuCDM}


\begin{table}
\centering
\setlength\tabcolsep{8pt}
\renewcommand\arraystretch{1.4}
\makebox[\textwidth][c]{
\begin{tabular}{c|c|c|c|c|c|c}
\toprule
Catalogs        & clean  &  dv-LRG &  dv-QSO-G  & dv-QSO-L  
                & dv-ELGcatas  &  dv-slitless\\
\hline \hline
$\sumnu$       & $< 0.78 \, \rm{eV}$ & $< 0.69 \, \rm{eV}$ & $< 1.40 \, \rm{eV}$  & $< 0.94 \, \rm{eV}$ 
                & $< 0.84 \, \rm{eV}$ & $< 0.85 \, \rm{eV}$  \\
\bottomrule
\end{tabular}}
\caption{Comparison of 95\% confidence upper bounds on the sum of neutrino masses, $\sumnu$. Results are obtained using Full-Modeling analysis with $\sumnu$ varied and fitting power spectrum monopole and quadrupole of 500 mocks with covariance rescaled to $V_{25}$. Both clean and contaminated mock catalogs are considered with $\sumnu = 0.1\,\mathrm{eV}$. We find that redshift uncertainties degrade the neutrino mass constraint up to $\sim80\%$ in QSO-like smearing cases, whereas ELG-like redshift catastrophics and slitless-like errors show negligible effect.} 
\label{tab:1D constrain on Mnu}
\end{table}


For the massive neutrino case, we use the \QUIJOTE $Mnu\_p$ catalogs with $\sumnu=0.1 \, \rm{eV}$ to build our clean and contaminated mock catalogs. During the fitting, we allow $\sumnu$ to vary with a flat prior of $\,\mathcal{U}(0.0, 5.0) \, \rm{eV}$ in the $\LCDM$ framework. The upper limits of $\sumnu$ at 95\% confidence level for all mocks analyses are presented in \cref{tab:1D constrain on Mnu}. We first note that using full-shape analysis alone to constrain the $\sumnu$ has slow convergence and large uncertainties, particularly for the low neutrino mass case. Nevertheless, we find that redshift uncertainty introduces extended tails in the $\sumnu$ posterior distribution, leading to a weaker constraint on the neutrino mass. Compared to the clean case ($<0.78 \, \rm{eV}$), the 95\% upper limits increase by up to $\sim 80\%$ when QSO Gaussian smearing is included. This degradation occurs because both massive neutrinos and redshift uncertainty have similar suppression on small-scale power spectrum. Massive neutrinos do so through their thermal velocities and free-streaming, while redshift uncertainty smears the clustering structure along the line of sight. This degeneracy complicates the convergence and leads to weaker constraints on neutrino mass. In contrast, we find no noticeable bias in the $\sumnu$ measurements due to ELG-like catastrophic failures or slitles-like redshift errors. 

\section{Conclusions}\label{sec:conclusions}

In this paper, we investigate the impact of spectroscopic redshift errors on cosmological measurements. \cref{sec:data spectroscopic survey and redshift errors} introduces the redshift errors, which can affect redshift measurements in both ground-based (fiber-fed) and space-based (slitless) spectroscopic surveys \cite{ross_completed_2020, yu_elg_2025, euclid_collaboration_euclid_2025-1}. There are two main types of redshift errors. Redshift uncertainty refer to small errors present in all redshift measurements, and redshift catastrophic failures correspond to large deviations from the true values in a small fraction of the measurements. Both type of errors distort the observed clustering pattern and can lead to biases in the inferred cosmological results.


The effect of redshift errors can be quantified and modeled using repeat observations. In \cref{sec:data contaminated mocks} we use the \QUIJOTE simulation suite to construct our contaminated mock catalogs. For redshift uncertainty, we consider Gaussian or Lorentzian profiles for LRG-like and QSO-like smearing~\cite{zarrouk_clustering_2018, yu_desi_2023}. For redshift catastrophics, we use a log-normal distribution~\cite{yu_elg_2025} with contamination rate $\fc=1\%$ to model ELG-like catastrophics. The slitless-like errors assume a hypothetical scenario combining $\fc=5\%$ catastrophics and LRG-like uncertainty to simulate the observing condition in slitless survey. The following \cref{sec:methods} describes the methodologies to perform the full-shape power spectrum analysis. In our baseline fitting, we use clustering models based on the effective field theory ($\EFT$) of LSS, without including the damping term or the multiplicative factor. We implement both ShapeFit and Full-Modeling fitting approach to extract cosmological information. The analysis is targeted on DESI- and \textit{Euclid}-like surveys with effective volume $V_{25}= 25\,\hgpc$. Our key findings are summarized below:

Redshift uncertainty introduces diffusive effect in the clustering, leading to a scale-dependent suppression to the observed power spectrum. The $\EFT$ cournterterms, parameterized by $\alpha_2$ and $\alpha_4$, capture this smearing effect, resulting in compressed and cosmological parameter estimates difference $<5\%$ compared to the uncontaminated case. In contrast, redshift catastrophics cause an overall reduction on the power spectrum amplitude that approximately scales as $(1 - \fc)^2$. While this effect is negligible (difference $<6\%$) for ELG-like catastrophics, it significantly biases the estimation of amplitude-related parameters for slitless-like errors, namely $df$ in ShapeFit and $\logA$ in the Full-Modeling. In that case, we observe shifts of $6\%{-}16\%$, corresponding to a $\sim2.2\sigma$ deviation relative to the clean case. To model this effect, we use the correction factor $(1 - \fc)^2$ applied to the power spectrum in \cref{eq:catastrophics simple model}. When freeing $\fc$, the extended model mitigates the bias in $\logA$ to below $0.1\sigma$, but leads to $67\%$ degradation in its precision due to the strong degeneracy between $\fc$ and $\logA$. Fixing $\fc$ to its expected value in the fit restores the constraining power with a modest $1.0\sigma$ deviation. This highlights the importance of accurately estimating the catastrophic failure rate $\fc$ and applying the appropriate correction model to ensure unbiased cosmological results in slitless spectroscopic surveys.

Beyond the vanilla $\Lambda$CDM model, we explore the impact of redshift errors on the measurement of evolving dark energy and massive neutrinos. For the $\wCDM$ model, we find that redshift errors do not bias the estimates of $w_0$ and $w_a$, although they can potentially enhance projection effects. In the massive neutrino case, redshift errors weaken the constraint on $\sumnu$ up to 80\% when considering QSO-like Gaussian smearing. 



For future work, we plan to apply our framework to the DESI DR2 analysis. The DESI DR2 dataset provides a large sample of repeat observations well-suited for characterizing redshift errors. Using this dataset, we aim to jointly model redshift uncertainty and catastrophic failures for each tracer population across a wide range of redshift. We also plan to conduct more detailed investigations of redshift errors, including lightcone effects, redshift evolution, impact on all various clustering statistics and so on.

\acknowledgments 
SH, AR, JY and JPK, acknowledge support from the Swiss National Science Foundation (SNF) `Cosmology with 3D Maps of the Universe'' research grant 200020\_207379. This research was essentially supported by computational resources provided by the High Performance Computing (HPC) facility of University of Geneva (UNIGE), especially for the Bamboo cluster.

\bibliographystyle{JHEP}
\bibliography{references.bib}

@article{damico_taming_2024,
	title = {Taming redshift-space distortion effects in the {EFTofLSS} and its application to data},
	volume = {2024},
	issn = {1475-7516},
	url = {http://arxiv.org/abs/2110.00016},
	doi = {10.1088/1475-7516/2024/01/037},
	number = {01},
	urldate = {2025-11-26},
	journal = {Journal of Cosmology and Astroparticle Physics},
	author = {D'Amico, Guido and Senatore, Leonardo and Zhang, Pierre and Nishimichi, Takahiro},
	month = jan,
	year = {2024},
	note = {arXiv:2110.00016 [astro-ph]},
	pages = {037},
}

@misc{senatore_redshift_2014,
	title = {Redshift {Space} {Distortions} in the {Effective} {Field} {Theory} of {Large} {Scale} {Structures}},
	url = {http://arxiv.org/abs/1409.1225},
	doi = {10.48550/arXiv.1409.1225},
	urldate = {2025-11-26},
	publisher = {arXiv},
	author = {Senatore, Leonardo and Zaldarriaga, Matias},
	month = sep,
	year = {2014},
	note = {arXiv:1409.1225 [astro-ph]},
}

@misc{desi_collaboration_desi_2016,
	title = {The {DESI} {Experiment} {Part} {I}: {Science},{Targeting}, and {Survey} {Design}},
	shorttitle = {The {DESI} {Experiment} {Part} {I}},
	url = {http://arxiv.org/abs/1611.00036},
	doi = {10.48550/arXiv.1611.00036},
	urldate = {2025-07-31},
	publisher = {arXiv},
	author = {{DESI Collaboration} and Aghamousa, Amir and Aguilar, Jessica and Ahlen, Steve and Alam, Shadab and Allen, Lori E. and Prieto, Carlos Allende and Annis, James and Bailey, Stephen and Balland, Christophe and Ballester, Otger and Baltay, Charles and Beaufore, Lucas and Bebek, Chris and Beers, Timothy C. and Bell, Eric F. and Bernal, José Luis and Besuner, Robert and Beutler, Florian and Blake, Chris and Bleuler, Hannes and Blomqvist, Michael and Blum, Robert and Bolton, Adam S. and Briceno, Cesar and Brooks, David and Brownstein, Joel R. and Buckley-Geer, Elizabeth and Burden, Angela and Burtin, Etienne and Busca, Nicolas G. and Cahn, Robert N. and Cai, Yan-Chuan and Cardiel-Sas, Laia and Carlberg, Raymond G. and Carton, Pierre-Henri and Casas, Ricard and Castander, Francisco J. and Cervantes-Cota, Jorge L. and Claybaugh, Todd M. and Close, Madeline and Coker, Carl T. and Cole, Shaun and Comparat, Johan and Cooper, Andrew P. and Cousinou, M.-C. and Crocce, Martin and Cuby, Jean-Gabriel and Cunningham, Daniel P. and Davis, Tamara M. and Dawson, Kyle S. and Macorra, Axel de la and Vicente, Juan De and Delubac, Timothée and Derwent, Mark and Dey, Arjun and Dhungana, Govinda and Ding, Zhejie and Doel, Peter and Duan, Yutong T. and Ealet, Anne and Edelstein, Jerry and Eftekharzadeh, Sarah and Eisenstein, Daniel J. and Elliott, Ann and Escoffier, Stéphanie and Evatt, Matthew and Fagrelius, Parker and Fan, Xiaohui and Fanning, Kevin and Farahi, Arya and Farihi, Jay and Favole, Ginevra and Feng, Yu and Fernandez, Enrique and Findlay, Joseph R. and Finkbeiner, Douglas P. and Fitzpatrick, Michael J. and Flaugher, Brenna and Flender, Samuel and Font-Ribera, Andreu and Forero-Romero, Jaime E. and Fosalba, Pablo and Frenk, Carlos S. and Fumagalli, Michele and Gaensicke, Boris T. and Gallo, Giuseppe and Garcia-Bellido, Juan and Gaztanaga, Enrique and Fusillo, Nicola Pietro Gentile and Gerard, Terry and Gershkovich, Irena and Giannantonio, Tommaso and Gillet, Denis and Gonzalez-de-Rivera, Guillermo and Gonzalez-Perez, Violeta and Gott, Shelby and Graur, Or and Gutierrez, Gaston and Guy, Julien and Habib, Salman and Heetderks, Henry and Heetderks, Ian and Heitmann, Katrin and Hellwing, Wojciech A. and Herrera, David A. and Ho, Shirley and Holland, Stephen and Honscheid, Klaus and Huff, Eric and Hutchinson, Timothy A. and Huterer, Dragan and Hwang, Ho Seong and Laguna, Joseph Maria Illa and Ishikawa, Yuzo and Jacobs, Dianna and Jeffrey, Niall and Jelinsky, Patrick and Jennings, Elise and Jiang, Linhua and Jimenez, Jorge and Johnson, Jennifer and Joyce, Richard and Jullo, Eric and Juneau, Stéphanie and Kama, Sami and Karcher, Armin and Karkar, Sonia and Kehoe, Robert and Kennamer, Noble and Kent, Stephen and Kilbinger, Martin and Kim, Alex G. and Kirkby, David and Kisner, Theodore and Kitanidis, Ellie and Kneib, Jean-Paul and Koposov, Sergey and Kovacs, Eve and Koyama, Kazuya and Kremin, Anthony and Kron, Richard and Kronig, Luzius and Kueter-Young, Andrea and Lacey, Cedric G. and Lafever, Robin and Lahav, Ofer and Lambert, Andrew and Lampton, Michael and Landriau, Martin and Lang, Dustin and Lauer, Tod R. and Goff, Jean-Marc Le and Guillou, Laurent Le and Suu, Auguste Le Van and Lee, Jae Hyeon and Lee, Su-Jeong and Leitner, Daniela and Lesser, Michael and Levi, Michael E. and L'Huillier, Benjamin and Li, Baojiu and Liang, Ming and Lin, Huan and Linder, Eric and Loebman, Sarah R. and Lukić, Zarija and Ma, Jun and MacCrann, Niall and Magneville, Christophe and Makarem, Laleh and Manera, Marc and Manser, Christopher J. and Marshall, Robert and Martini, Paul and Massey, Richard and Matheson, Thomas and McCauley, Jeremy and McDonald, Patrick and McGreer, Ian D. and Meisner, Aaron and Metcalfe, Nigel and Miller, Timothy N. and Miquel, Ramon and Moustakas, John and Myers, Adam and Naik, Milind and Newman, Jeffrey A. and Nichol, Robert C. and Nicola, Andrina and Costa, Luiz Nicolati da and Nie, Jundan and Niz, Gustavo and Norberg, Peder and Nord, Brian and Norman, Dara and Nugent, Peter and O'Brien, Thomas and Oh, Minji and Olsen, Knut A. G. and Padilla, Cristobal and Padmanabhan, Hamsa and Padmanabhan, Nikhil and Palanque-Delabrouille, Nathalie and Palmese, Antonella and Pappalardo, Daniel and Pâris, Isabelle and Park, Changbom and Patej, Anna and Peacock, John A. and Peiris, Hiranya V. and Peng, Xiyan and Percival, Will J. and Perruchot, Sandrine and Pieri, Matthew M. and Pogge, Richard and Pollack, Jennifer E. and Poppett, Claire and Prada, Francisco and Prakash, Abhishek and Probst, Ronald G. and Rabinowitz, David and Raichoor, Anand and Ree, Chang Hee and Refregier, Alexandre and Regal, Xavier and Reid, Beth and Reil, Kevin and Rezaie, Mehdi and Rockosi, Constance M. and Roe, Natalie and Ronayette, Samuel and Roodman, Aaron and Ross, Ashley J. and Ross, Nicholas P. and Rossi, Graziano and Rozo, Eduardo and Ruhlmann-Kleider, Vanina and Rykoff, Eli S. and Sabiu, Cristiano and Samushia, Lado and Sanchez, Eusebio and Sanchez, Javier and Schlegel, David J. and Schneider, Michael and Schubnell, Michael and Secroun, Aurélia and Seljak, Uros and Seo, Hee-Jong and Serrano, Santiago and Shafieloo, Arman and Shan, Huanyuan and Sharples, Ray and Sholl, Michael J. and Shourt, William V. and Silber, Joseph H. and Silva, David R. and Sirk, Martin M. and Slosar, Anze and Smith, Alex and Smoot, George F. and Som, Debopam and Song, Yong-Seon and Sprayberry, David and Staten, Ryan and Stefanik, Andy and Tarle, Gregory and Tie, Suk Sien and Tinker, Jeremy L. and Tojeiro, Rita and Valdes, Francisco and Valenzuela, Octavio and Valluri, Monica and Vargas-Magana, Mariana and Verde, Licia and Walker, Alistair R. and Wang, Jiali and Wang, Yuting and Weaver, Benjamin A. and Weaverdyck, Curtis and Wechsler, Risa H. and Weinberg, David H. and White, Martin and Yang, Qian and Yeche, Christophe and Zhang, Tianmeng and Zhao, Gong-Bo and Zheng, Yi and Zhou, Xu and Zhou, Zhimin and Zhu, Yaling and Zou, Hu and Zu, Ying},
	month = dec,
	year = {2016},
	note = {arXiv:1611.00036 [astro-ph]},
}

@misc{csst_collaboration_introduction_2025,
	title = {Introduction to the {China} {Space} {Station} {Telescope} ({CSST})},
	url = {http://arxiv.org/abs/2507.04618},
	doi = {10.48550/arXiv.2507.04618},
	urldate = {2025-08-02},
	publisher = {arXiv},
	author = {{CSST Collaboration} and Gong, Yan and Miao, Haitao and Zhan, Hu and Li, Zhao-Yu and Shangguan, Jinyi and Li, Haining and Liu, Chao and Chen, Xuefei and Yuan, Haibo and Zhou, Jilin and Liu, Hui-Gen and Yu, Cong and Ji, Jianghui and Qi, Zhaoxiang and Liu, Jiacheng and Dai, Zigao and Wang, Xiaofeng and Zheng, Zhenya and Hao, Lei and Dou, Jiangpei and Ao, Yiping and Lin, Zhenhui and Zhang, Kun and Wang, Wei and Sun, Guotong and Li, Ran and Li, Guoliang and Xu, Youhua and Li, Xinfeng and Li, Shengyang and Wu, Peng and Zhang, Jiuxing and Wang, Bo and Bai, Jinming and Cai, Yi-Fu and Cai, Zheng and Chan, Kwan Chuen and Chang, Jin and Chen, Xiaodian and Chen, Xuelei and Chen, Yuqin and Chen, Yun and Cui, Wei and Du, Pu and Duan, Wenying and Fan, Junhui and Fan, LuLu and Fan, Zhou and Fan, Zuhui and Fang, Taotao and Fu, Jianning and Fu, Liping and Fu, Zhensen and Gao, Jian and Gu, Shenghong and Gu, Yidong and Guo, Qi and Han, Zhanwen and Huang, Zhiqi and Ho, Luis C. and Jiang, Linhua and Jing, Yipeng and Kang, Xi and Kong, Xu and Li, Chengyuan and Li, Di and Li, Jing and Li, Nan and Li, Yang A. and Liao, Shilong and Lin, Weipeng and Liu, Fengshan and Liu, Jifeng and Liu, Xiangkun and Mao, Ruiqing and Mao, Shude and Meng, Xianmin and Pang, Xiaoying and Peng, Xiyan and Peng, Yingjie and Shan, Huanyuan and Shen, Juntai and Shen, Shiyin and Shen, Zhiqiang and Shi, Sheng-Cai and Shi, Yong and Tan, Siyuan and Tian, Hao and Wang, Jianmin and Wang, Jun-Xian and Wang, Xin and Wang, Yuting and Wu, Hong and Wu, Jingwen and Wu, Xuebing and Xu, Chun and Xue, Xiang-Xiang and Xue, Yongquan and Yang, Ji and Yang, Xiaohu and Yao, Qijun and Yuan, Fangting and Yuan, Zhen and Zhang, Jun and Zhang, Wei and Zhang, Xin and Zhao, Gang and Zhao, Gongbo and Zhong, Hongen and Zhong, Jing and Zhou, Liyong and Zu, Ying},
	month = jul,
	year = {2025},
	note = {arXiv:2507.04618 [astro-ph]},
}

@misc{desi_collaboration_desi_2025-1,
	title = {{DESI} 2024 {V}: {Full}-{Shape} {Galaxy} {Clustering} from {Galaxies} and {Quasars}},
	shorttitle = {{DESI} 2024 {V}},
	url = {http://arxiv.org/abs/2411.12021},
	doi = {10.48550/arXiv.2411.12021},
	urldate = {2025-07-31},
	publisher = {arXiv},
	author = {{DESI Collaboration} and Adame, A. G. and Aguilar, J. and Ahlen, S. and Alam, S. and Alexander, D. M. and Alvarez, M. and Alves, O. and Anand, A. and Andrade, U. and Armengaud, E. and Avila, S. and Aviles, A. and Awan, H. and Bailey, S. and Baltay, C. and Bault, A. and Behera, J. and BenZvi, S. and Beutler, F. and Bianchi, D. and Blake, C. and Blum, R. and Brieden, S. and Brodzeller, A. and Brooks, D. and Buckley-Geer, E. and Burtin, E. and Calderon, R. and Canning, R. and Rosell, A. Carnero and Cereskaite, R. and Cervantes-Cota, J. L. and Chabanier, S. and Chaussidon, E. and Chaves-Montero, J. and Chen, S. and Chen, X. and Claybaugh, T. and Cole, S. and Cuceu, A. and Davis, T. M. and Dawson, K. and Macorra, A. de la and Mattia, A. de and Deiosso, N. and Dey, A. and Dey, B. and Ding, Z. and Doel, P. and Edelstein, J. and Eftekharzadeh, S. and Eisenstein, D. J. and Elliott, A. and Fagrelius, P. and Fanning, K. and Ferraro, S. and Ereza, J. and Findlay, N. and Flaugher, B. and Font-Ribera, A. and Forero-Sánchez, D. and Forero-Romero, J. E. and Garcia-Quintero, C. and Garrison, L. H. and Gaztañaga, E. and Gil-Marín, H. and Gontcho, S. Gontcho A. and Gonzalez-Morales, A. X. and Gonzalez-Perez, V. and Gordon, C. and Green, D. and Gruen, D. and Gsponer, R. and Gutierrez, G. and Guy, J. and Hadzhiyska, B. and Hahn, C. and Hanif, M. M. S. and Herrera-Alcantar, H. K. and Honscheid, K. and Howlett, C. and Huterer, D. and Iršič, V. and Ishak, M. and Juneau, S. and Karaçaylı, N. G. and Kehoe, R. and Kent, S. and Kirkby, D. and Kong, H. and Koposov, S. E. and Kremin, A. and Krolewski, A. and Lai, Y. and Lan, T.-W. and Landriau, M. and Lang, D. and Lasker, J. and Goff, J. M. Le and Guillou, L. Le and Leauthaud, A. and Levi, M. E. and Li, T. S. and Lodha, K. and Magneville, C. and Manera, M. and Margala, D. and Martini, P. and Maus, M. and McDonald, P. and Medina-Varela, L. and Meisner, A. and Mena-Fernández, J. and Miquel, R. and Moon, J. and Moore, S. and Moustakas, J. and Mueller, E. and Muñoz-Gutiérrez, A. and Myers, A. D. and Nadathur, S. and Napolitano, L. and Neveux, R. and Newman, J. A. and Nguyen, N. M. and Nie, J. and Niz, G. and Noriega, H. E. and Padmanabhan, N. and Paillas, E. and Palanque-Delabrouille, N. and Pan, J. and Penmetsa, S. and Percival, W. J. and Pieri, M. M. and Pinon, M. and Poppett, C. and Porredon, A. and Prada, F. and Pérez-Fernández, A. and Pérez-Ràfols, I. and Rabinowitz, D. and Raichoor, A. and Ramírez-Pérez, C. and Ramirez-Solano, S. and Rashkovetskyi, M. and Ravoux, C. and Rezaie, M. and Rich, J. and Rocher, A. and Rockosi, C. and Rodríguez-Martínez, F. and Roe, N. A. and Rosado-Marin, A. and Ross, A. J. and Rossi, G. and Ruggeri, R. and Ruhlmann-Kleider, V. and Samushia, L. and Sanchez, E. and Saulder, C. and Schlafly, E. F. and Schlegel, D. and Schubnell, M. and Seo, H. and Sharples, R. and Silber, J. and Slosar, A. and Smith, A. and Sprayberry, D. and Tan, T. and Tarlé, G. and Trusov, S. and Vaisakh, R. and Valcin, D. and Valdes, F. and Vargas-Magaña, M. and Verde, L. and Walther, M. and Wang, B. and Wang, M. S. and Weaver, B. A. and Weaverdyck, N. and Wechsler, R. H. and Weinberg, D. H. and White, M. and Wilson, M. J. and Yu, J. and Yu, Y. and Yuan, S. and Yèche, C. and Zaborowski, E. A. and Zarrouk, P. and Zhang, H. and Zhao, C. and Zhao, R. and Zhou, R. and Zou, H.},
	month = mar,
	year = {2025},
	note = {arXiv:2411.12021 [astro-ph]},
}

@article{desi_collaboration_desi_2025-2,
	title = {{DESI} 2024 {III}: {Baryon} {Acoustic} {Oscillations} from {Galaxies} and {Quasars}},
	volume = {2025},
	issn = {1475-7516},
	shorttitle = {{DESI} 2024 {III}},
	url = {http://arxiv.org/abs/2404.03000},
	doi = {10.1088/1475-7516/2025/04/012},
	number = {04},
	urldate = {2025-07-31},
	journal = {Journal of Cosmology and Astroparticle Physics},
	author = {{DESI Collaboration} and Adame, A. G. and Aguilar, J. and Ahlen, S. and Alam, S. and Alexander, D. M. and Alvarez, M. and Alves, O. and Anand, A. and Andrade, U. and Armengaud, E. and Avila, S. and Aviles, A. and Awan, H. and Bailey, S. and Baltay, C. and Bault, A. and Behera, J. and BenZvi, S. and Beutler, F. and Bianchi, D. and Blake, C. and Blum, R. and Brieden, S. and Brodzeller, A. and Brooks, D. and Buckley-Geer, E. and Burtin, E. and Calderon, R. and Canning, R. and Rosell, A. Carnero and Cereskaite, R. and Cervantes-Cota, J. L. and Chabanier, S. and Chaussidon, E. and Chaves-Montero, J. and Chen, S. and Chen, X. and Claybaugh, T. and Cole, S. and Cuceu, A. and Davis, T. M. and Dawson, K. and Macorra, A. de la and Mattia, A. de and Deiosso, N. and Dey, A. and Dey, B. and Ding, Z. and Doel, P. and Edelstein, J. and Eftekharzadeh, S. and Eisenstein, D. J. and Elliott, A. and Fagrelius, P. and Fanning, K. and Ferraro, S. and Ereza, J. and Findlay, N. and Flaugher, B. and Font-Ribera, A. and Forero-Sánchez, D. and Forero-Romero, J. E. and Garcia-Quintero, C. and Gaztañaga, E. and Gil-Marín, H. and Gontcho, S. Gontcho A. and Gonzalez-Morales, A. X. and Gonzalez-Perez, V. and Gordon, C. and Green, D. and Gruen, D. and Gsponer, R. and Gutierrez, G. and Guy, J. and Hadzhiyska, B. and Hahn, C. and Hanif, M. M. S. and Herrera-Alcantar, H. K. and Honscheid, K. and Howlett, C. and Huterer, D. and Iršič, V. and Ishak, M. and Juneau, S. and Karaçaylı, N. G. and Kehoe, R. and Kent, S. and Kirkby, D. and Kremin, A. and Krolewski, A. and Lai, Y. and Lan, T.-W. and Landriau, M. and Lang, D. and Lasker, J. and Goff, J. M. Le and Guillou, L. Le and Leauthaud, A. and Levi, M. E. and Li, T. S. and Linder, E. and Lodha, K. and Magneville, C. and Manera, M. and Margala, D. and Martini, P. and Maus, M. and McDonald, P. and Medina-Varela, L. and Meisner, A. and Mena-Fernández, J. and Miquel, R. and Moon, J. and Moore, S. and Moustakas, J. and Mudur, N. and Mueller, E. and Muñoz-Gutiérrez, A. and Myers, A. D. and Nadathur, S. and Napolitano, L. and Neveux, R. and Newman, J. A. and Nguyen, N. M. and Nie, J. and Niz, G. and Noriega, H. E. and Padmanabhan, N. and Paillas, E. and Palanque-Delabrouille, N. and Pan, J. and Penmetsa, S. and Percival, W. J. and Pieri, M. and Pinon, M. and Poppett, C. and Porredon, A. and Prada, F. and Pérez-Fernández, A. and Pérez-Ràfols, I. and Rabinowitz, D. and Raichoor, A. and Ramírez-Pérez, C. and Ramirez-Solano, S. and Rashkovetskyi, M. and Rezaie, M. and Rich, J. and Rocher, A. and Rockosi, C. and Roe, N. A. and Rosado-Marin, A. and Ross, A. J. and Rossi, G. and Ruggeri, R. and Ruhlmann-Kleider, V. and Samushia, L. and Sanchez, E. and Saulder, C. and Schlafly, E. F. and Schlegel, D. and Schubnell, M. and Seo, H. and Sharples, R. and Silber, J. and Slosar, A. and Smith, A. and Sprayberry, D. and Swanson, J. and Tan, T. and Tarlé, G. and Trusov, S. and Vaisakh, R. and Valcin, D. and Valdes, F. and Vargas-Magaña, M. and Verde, L. and Walther, M. and Wang, B. and Wang, M. S. and Weaver, B. A. and Weaverdyck, N. and Wechsler, R. H. and Weinberg, D. H. and White, M. and Yu, J. and Yu, Y. and Yuan, S. and Yèche, C. and Zaborowski, E. A. and Zarrouk, P. and Zhang, H. and Zhao, C. and Zhao, R. and Zhou, R. and Zou, H.},
	month = apr,
	year = {2025},
	pages = {012},
}

@article{desi_collaboration_desi_2025-3,
	title = {{DESI} 2024 {VII}: {Cosmological} {Constraints} from the {Full}-{Shape} {Modeling} of {Clustering} {Measurements}},
	volume = {2025},
	issn = {1475-7516},
	shorttitle = {{DESI} 2024 {VII}},
	url = {http://arxiv.org/abs/2411.12022},
	doi = {10.1088/1475-7516/2025/07/028},
	number = {07},
	urldate = {2025-07-31},
	journal = {Journal of Cosmology and Astroparticle Physics},
	author = {{DESI Collaboration} and Adame, A. G. and Aguilar, J. and Ahlen, S. and Alam, S. and Alexander, D. M. and Prieto, C. Allende and Alvarez, M. and Alves, O. and Anand, A. and Andrade, U. and Armengaud, E. and Avila, S. and Aviles, A. and Awan, H. and Bahr-Kalus, B. and Bailey, S. and Baltay, C. and Bault, A. and Behera, J. and BenZvi, S. and Beutler, F. and Bianchi, D. and Blake, C. and Blum, R. and Bonici, M. and Brieden, S. and Brodzeller, A. and Brooks, D. and Buckley-Geer, E. and Burtin, E. and Calderon, R. and Canning, R. and Rosell, A. Carnero and Cereskaite, R. and Cervantes-Cota, J. L. and Chabanier, S. and Chaussidon, E. and Chaves-Montero, J. and Chebat, D. and Chen, S. and Chen, X. and Claybaugh, T. and Cole, S. and Cuceu, A. and Davis, T. M. and Dawson, K. and Macorra, A. de la and Mattia, A. de and Deiosso, N. and Dey, A. and Dey, B. and Ding, Z. and Doel, P. and Edelstein, J. and Eftekharzadeh, S. and Eisenstein, D. J. and Elbers, W. and Elliott, A. and Fagrelius, P. and Fanning, K. and Ferraro, S. and Ereza, J. and Findlay, N. and Flaugher, B. and Font-Ribera, A. and Forero-Sánchez, D. and Forero-Romero, J. E. and Frenk, C. S. and Garcia-Quintero, C. and Garrison, L. H. and Gaztañaga, E. and Gil-Marín, H. and Gontcho, S. Gontcho A. and Gonzalez-Morales, A. X. and Gonzalez-Perez, V. and Gordon, C. and Green, D. and Gruen, D. and Gsponer, R. and Gutierrez, G. and Guy, J. and Hadzhiyska, B. and Hahn, C. and Hanif, M. M. S. and Herrera-Alcantar, H. K. and Honscheid, K. and Howlett, C. and Huterer, D. and Iršič, V. and Ishak, M. and Joyce, R. and Juneau, S. and Karaçaylı, N. G. and Kehoe, R. and Kent, S. and Kirkby, D. and Kong, H. and Koposov, S. E. and Kremin, A. and Krolewski, A. and Lahav, O. and Lai, Y. and Lan, T.-W. and Landriau, M. and Lang, D. and Lasker, J. and Goff, J. M. Le and Guillou, L. Le and Leauthaud, A. and Levi, M. E. and Li, T. S. and Lodha, K. and Magneville, C. and Manera, M. and Margala, D. and Martini, P. and Matthewson, W. and Maus, M. and McDonald, P. and Medina-Varela, L. and Meisner, A. and Mena-Fernández, J. and Miquel, R. and Moon, J. and Moore, S. and Moustakas, J. and Mudur, N. and Mueller, E. and Muñoz-Gutiérrez, A. and Myers, A. D. and Nadathur, S. and Napolitano, L. and Neveux, R. and Newman, J. A. and Nguyen, N. M. and Nie, J. and Niz, G. and Noriega, H. E. and Padmanabhan, N. and Paillas, E. and Palanque-Delabrouille, N. and Pan, J. and Penmetsa, S. and Percival, W. J. and Pieri, M. M. and Pinon, M. and Poppett, C. and Porredon, A. and Prada, F. and Pérez-Fernández, A. and Pérez-Ràfols, I. and Rabinowitz, D. and Raichoor, A. and Ramírez-Pérez, C. and Ramirez-Solano, S. and Rashkovetskyi, M. and Ravoux, C. and Rezaie, M. and Rich, J. and Rocher, A. and Rockosi, C. and Roe, N. A. and Rosado-Marin, A. and Ross, A. J. and Rossi, G. and Ruggeri, R. and Ruhlmann-Kleider, V. and Samushia, L. and Sanchez, E. and Saulder, C. and Schlafly, E. F. and Schlegel, D. and Schubnell, M. and Seo, H. and Shafieloo, A. and Sharples, R. and Silber, J. and Slosar, A. and Smith, A. and Sprayberry, D. and Tan, T. and Tarlé, G. and Taylor, P. and Trusov, S. and Vaisakh, R. and Valcin, D. and Valdes, F. and Valogiannis, G. and Vargas-Magaña, M. and Verde, L. and Walther, M. and Wang, B. and Wang, M. S. and Weaver, B. A. and Weaverdyck, N. and Wechsler, R. H. and Weinberg, D. H. and White, M. and Wilson, M. J. and Yi, L. and Yu, J. and Yu, Y. and Yuan, S. and Yèche, C. and Zaborowski, E. A. and Zarrouk, P. and Zhang, H. and Zhao, C. and Zhao, R. and Zhou, R. and Zhuang, T. and Zou, H.},
	month = jul,
	year = {2025},
	note = {arXiv:2411.12022 [astro-ph]},
	pages = {028},
}

@misc{desi_collaboration_desi_2016-1,
	title = {The {DESI} {Experiment} {Part} {II}: {Instrument} {Design}},
	shorttitle = {The {DESI} {Experiment} {Part} {II}},
	url = {http://arxiv.org/abs/1611.00037},
	doi = {10.48550/arXiv.1611.00037},
	urldate = {2025-07-31},
	publisher = {arXiv},
	author = {{DESI Collaboration} and Aghamousa, Amir and Aguilar, Jessica and Ahlen, Steve and Alam, Shadab and Allen, Lori E. and Prieto, Carlos Allende and Annis, James and Bailey, Stephen and Balland, Christophe and Ballester, Otger and Baltay, Charles and Beaufore, Lucas and Bebek, Chris and Beers, Timothy C. and Bell, Eric F. and Bernal, José Luis and Besuner, Robert and Beutler, Florian and Blake, Chris and Bleuler, Hannes and Blomqvist, Michael and Blum, Robert and Bolton, Adam S. and Briceno, Cesar and Brooks, David and Brownstein, Joel R. and Buckley-Geer, Elizabeth and Burden, Angela and Burtin, Etienne and Busca, Nicolas G. and Cahn, Robert N. and Cai, Yan-Chuan and Cardiel-Sas, Laia and Carlberg, Raymond G. and Carton, Pierre-Henri and Casas, Ricard and Castander, Francisco J. and Cervantes-Cota, Jorge L. and Claybaugh, Todd M. and Close, Madeline and Coker, Carl T. and Cole, Shaun and Comparat, Johan and Cooper, Andrew P. and Cousinou, M.-C. and Crocce, Martin and Cuby, Jean-Gabriel and Cunningham, Daniel P. and Davis, Tamara M. and Dawson, Kyle S. and Macorra, Axel de la and Vicente, Juan De and Delubac, Timothée and Derwent, Mark and Dey, Arjun and Dhungana, Govinda and Ding, Zhejie and Doel, Peter and Duan, Yutong T. and Ealet, Anne and Edelstein, Jerry and Eftekharzadeh, Sarah and Eisenstein, Daniel J. and Elliott, Ann and Escoffier, Stéphanie and Evatt, Matthew and Fagrelius, Parker and Fan, Xiaohui and Fanning, Kevin and Farahi, Arya and Farihi, Jay and Favole, Ginevra and Feng, Yu and Fernandez, Enrique and Findlay, Joseph R. and Finkbeiner, Douglas P. and Fitzpatrick, Michael J. and Flaugher, Brenna and Flender, Samuel and Font-Ribera, Andreu and Forero-Romero, Jaime E. and Fosalba, Pablo and Frenk, Carlos S. and Fumagalli, Michele and Gaensicke, Boris T. and Gallo, Giuseppe and Garcia-Bellido, Juan and Gaztanaga, Enrique and Fusillo, Nicola Pietro Gentile and Gerard, Terry and Gershkovich, Irena and Giannantonio, Tommaso and Gillet, Denis and Gonzalez-de-Rivera, Guillermo and Gonzalez-Perez, Violeta and Gott, Shelby and Graur, Or and Gutierrez, Gaston and Guy, Julien and Habib, Salman and Heetderks, Henry and Heetderks, Ian and Heitmann, Katrin and Hellwing, Wojciech A. and Herrera, David A. and Ho, Shirley and Holland, Stephen and Honscheid, Klaus and Huff, Eric and Hutchinson, Timothy A. and Huterer, Dragan and Hwang, Ho Seong and Laguna, Joseph Maria Illa and Ishikawa, Yuzo and Jacobs, Dianna and Jeffrey, Niall and Jelinsky, Patrick and Jennings, Elise and Jiang, Linhua and Jimenez, Jorge and Johnson, Jennifer and Joyce, Richard and Jullo, Eric and Juneau, Stéphanie and Kama, Sami and Karcher, Armin and Karkar, Sonia and Kehoe, Robert and Kennamer, Noble and Kent, Stephen and Kilbinger, Martin and Kim, Alex G. and Kirkby, David and Kisner, Theodore and Kitanidis, Ellie and Kneib, Jean-Paul and Koposov, Sergey and Kovacs, Eve and Koyama, Kazuya and Kremin, Anthony and Kron, Richard and Kronig, Luzius and Kueter-Young, Andrea and Lacey, Cedric G. and Lafever, Robin and Lahav, Ofer and Lambert, Andrew and Lampton, Michael and Landriau, Martin and Lang, Dustin and Lauer, Tod R. and Goff, Jean-Marc Le and Guillou, Laurent Le and Suu, Auguste Le Van and Lee, Jae Hyeon and Lee, Su-Jeong and Leitner, Daniela and Lesser, Michael and Levi, Michael E. and L'Huillier, Benjamin and Li, Baojiu and Liang, Ming and Lin, Huan and Linder, Eric and Loebman, Sarah R. and Lukić, Zarija and Ma, Jun and MacCrann, Niall and Magneville, Christophe and Makarem, Laleh and Manera, Marc and Manser, Christopher J. and Marshall, Robert and Martini, Paul and Massey, Richard and Matheson, Thomas and McCauley, Jeremy and McDonald, Patrick and McGreer, Ian D. and Meisner, Aaron and Metcalfe, Nigel and Miller, Timothy N. and Miquel, Ramon and Moustakas, John and Myers, Adam and Naik, Milind and Newman, Jeffrey A. and Nichol, Robert C. and Nicola, Andrina and Costa, Luiz Nicolati da and Nie, Jundan and Niz, Gustavo and Norberg, Peder and Nord, Brian and Norman, Dara and Nugent, Peter and O'Brien, Thomas and Oh, Minji and Olsen, Knut A. G. and Padilla, Cristobal and Padmanabhan, Hamsa and Padmanabhan, Nikhil and Palanque-Delabrouille, Nathalie and Palmese, Antonella and Pappalardo, Daniel and Pâris, Isabelle and Park, Changbom and Patej, Anna and Peacock, John A. and Peiris, Hiranya V. and Peng, Xiyan and Percival, Will J. and Perruchot, Sandrine and Pieri, Matthew M. and Pogge, Richard and Pollack, Jennifer E. and Poppett, Claire and Prada, Francisco and Prakash, Abhishek and Probst, Ronald G. and Rabinowitz, David and Raichoor, Anand and Ree, Chang Hee and Refregier, Alexandre and Regal, Xavier and Reid, Beth and Reil, Kevin and Rezaie, Mehdi and Rockosi, Constance M. and Roe, Natalie and Ronayette, Samuel and Roodman, Aaron and Ross, Ashley J. and Ross, Nicholas P. and Rossi, Graziano and Rozo, Eduardo and Ruhlmann-Kleider, Vanina and Rykoff, Eli S. and Sabiu, Cristiano and Samushia, Lado and Sanchez, Eusebio and Sanchez, Javier and Schlegel, David J. and Schneider, Michael and Schubnell, Michael and Secroun, Aurélia and Seljak, Uros and Seo, Hee-Jong and Serrano, Santiago and Shafieloo, Arman and Shan, Huanyuan and Sharples, Ray and Sholl, Michael J. and Shourt, William V. and Silber, Joseph H. and Silva, David R. and Sirk, Martin M. and Slosar, Anze and Smith, Alex and Smoot, George F. and Som, Debopam and Song, Yong-Seon and Sprayberry, David and Staten, Ryan and Stefanik, Andy and Tarle, Gregory and Tie, Suk Sien and Tinker, Jeremy L. and Tojeiro, Rita and Valdes, Francisco and Valenzuela, Octavio and Valluri, Monica and Vargas-Magana, Mariana and Verde, Licia and Walker, Alistair R. and Wang, Jiali and Wang, Yuting and Weaver, Benjamin A. and Weaverdyck, Curtis and Wechsler, Risa H. and Weinberg, David H. and White, Martin and Yang, Qian and Yeche, Christophe and Zhang, Tianmeng and Zhao, Gong-Bo and Zheng, Yi and Zhou, Xu and Zhou, Zhimin and Zhu, Yaling and Zou, Hu and Zu, Ying},
	month = dec,
	year = {2016},
	note = {arXiv:1611.00037 [astro-ph]},
}

@misc{euclid_collaboration_euclid_2025,
	title = {Euclid {Quick} {Data} {Release} ({Q1}) -- {Data} release overview},
	url = {http://arxiv.org/abs/2503.15302},
	doi = {10.48550/arXiv.2503.15302},
	urldate = {2025-08-02},
	publisher = {arXiv},
	author = {{Euclid Collaboration} and Aussel, H. and Tereno, I. and Schirmer, M. and Alguero, G. and Altieri, B. and Balbinot, E. and Boer, T. de and Casenove, P. and Corcho-Caballero, P. and Furusawa, H. and Furusawa, J. and Hudson, M. J. and Jahnke, K. and Libet, G. and Macias-Perez, J. and Masoumzadeh, N. and Mohr, J. J. and Odier, J. and Scott, D. and Vassallo, T. and Kleijn, G. Verdoes and Zacchei, A. and Aghanim, N. and Amara, A. and Andreon, S. and Auricchio, N. and Awan, S. and Azzollini, R. and Baccigalupi, C. and Baldi, M. and Balestra, A. and Bardelli, S. and Basset, A. and Battaglia, P. and Belikov, A. N. and Bender, R. and Biviano, A. and Bonchi, A. and Bonino, D. and Branchini, E. and Brescia, M. and Brinchmann, J. and Camera, S. and Cañas-Herrera, G. and Capobianco, V. and Carbone, C. and Cardone, V. F. and Carretero, J. and Casas, S. and Castander, F. J. and Castellano, M. and Castignani, G. and Cavuoti, S. and Chambers, K. C. and Cimatti, A. and Colodro-Conde, C. and Congedo, G. and Conselice, C. J. and Conversi, L. and Copin, Y. and Courbin, F. and Courtois, H. M. and Cropper, M. and Cuby, J.-G. and Silva, A. Da and Silva, R. da and Degaudenzi, H. and Jong, J. T. A. de and Lucia, G. De and Giorgio, A. M. Di and Dinis, J. and Dolding, C. and Dole, H. and Douspis, M. and Dubath, F. and Duncan, C. A. J. and Dupac, X. and Dusini, S. and Ealet, A. and Escoffier, S. and Fabricius, M. and Farina, M. and Farinelli, R. and Faustini, F. and Ferriol, S. and Fotopoulou, S. and Fourmanoit, N. and Frailis, M. and Franceschi, E. and Franzetti, P. and Galeotta, S. and George, K. and Gillard, W. and Gillis, B. and Giocoli, C. and Gómez-Alvarez, P. and Gracia-Carpio, J. and Granett, B. R. and Grazian, A. and Grupp, F. and Guzzo, L. and Gwyn, S. and Haugan, S. V. H. and Herent, O. and Hoar, J. and Hoekstra, H. and Holliman, M. S. and Holmes, W. and Hook, I. M. and Hormuth, F. and Hornstrup, A. and Hudelot, P. and Ilić, S. and Jhabvala, M. and Joachimi, B. and Keihänen, E. and Kermiche, S. and Kiessling, A. and Kubik, B. and Kuijken, K. and Kümmel, M. and Kunz, M. and Kurki-Suonio, H. and Lahav, O. and Boulc'h, Q. Le and Brun, A. M. C. Le and Mignant, D. Le and Liebing, P. and Ligori, S. and Lilje, P. B. and Lindholm, V. and Lloro, I. and Mainetti, G. and Maino, D. and Maiorano, E. and Mansutti, O. and Marcin, S. and Marggraf, O. and Markovic, K. and Martinelli, M. and Martinet, N. and Marulli, F. and Massey, R. and Maurogordato, S. and McCracken, H. J. and Medinaceli, E. and Mei, S. and Melchior, M. and Mellier, Y. and Meneghetti, M. and Merlin, E. and Meylan, G. and Mora, A. and Moresco, M. and Morris, P. W. and Moscardini, L. and Mourre, S. and Nakajima, R. and Neissner, C. and Nichol, R. C. and Niemi, S.-M. and Nightingale, J. W. and Nutma, T. and Padilla, C. and Paltani, S. and Pasian, F. and Peacock, J. A. and Pedersen, K. and Percival, W. J. and Pettorino, V. and Pires, S. and Polenta, G. and Pollack, J. E. and Poncet, M. and Popa, L. A. and Pozzetti, L. and Racca, G. D. and Raison, F. and Rebolo, R. and Renzi, A. and Rhodes, J. and Riccio, G. and Rix, H.-W. and Romelli, E. and Roncarelli, M. and Rossetti, E. and Rusholme, B. and Saglia, R. and Sakr, Z. and Sánchez, A. G. and Sapone, D. and Sartoris, B. and Sauvage, M. and Schewtschenko, J. A. and Schneider, P. and Scodeggio, M. and Secroun, A. and Sefusatti, E. and Seidel, G. and Seiffert, M. and Serrano, S. and Simon, P. and Sirignano, C. and Sirri, G. and Skottfelt, J. and Mancini, A. Spurio and Stanco, L. and Steinwagner, J. and Surace, C. and Tallada-Crespí, P. and Tavagnacco, D. and Taylor, A. N. and Teplitz, H. I. and Tessore, N. and Toft, S. and Toledo-Moreo, R. and Torradeflot, F. and Tsyganov, A. and Tutusaus, I. and Valentijn, E. A. and Valenziano, L. and Valiviita, J. and Veropalumbo, A. and Wang, Y. and Weller, J. and Williams, O. R. and Zamorani, G. and Zerbi, F. M. and Zucca, E. and Allevato, V. and Ballardini, M. and Blake, R. P. and Bolzonella, M. and Bozzo, E. and Burigana, C. and Cabanac, R. and Calabrese, M. and Cappi, A. and Ferdinando, D. Di and Vigo, J. A. Escartin and Gabarra, L. and Hartley, W. G. and Huertas-Company, M. and Martín-Fleitas, J. and Matthew, S. and Maturi, M. and Mauri, N. and Metcalf, R. B. and Pezzotta, A. and Pöntinen, M. and Porciani, C. and Risso, I. and Scottez, V. and Sereno, M. and Tenti, M. and Viel, M. and Wiesmann, M. and Akrami, Y. and Alvi, S. and Andika, I. T. and Anselmi, S. and Archidiacono, M. and Atrio-Barandela, F. and Avila, S. and Bergamini, P. and Bertacca, D. and Bethermin, M. and Bisigello, L. and Blanchard, A. and Blot, L. and Böhringer, H. and Borgani, S. and Borlaff, A. S. and Brown, M. L. and Bruton, S. and Buitrago, F. and Calabro, A. and Calderone, G. and Quevedo, B. Camacho and Caro, F. and Carvalho, C. S. and Castro, T. and Charles, Y. and Cogato, F. and Conseil, S. and Cooray, A. R. and Costanzi, M. and Cucciati, O. and Davini, S. and Paolis, F. De and Desprez, G. and Díaz-Sánchez, A. and Diaz, J. J. and Domizio, S. Di and Diego, J. M. and Dimauro, P. and Duc, P.-A. and Enia, A. and Fang, Y. and Ferguson, A. M. N. and Ferrari, A. G. and Finoguenov, A. and Fontana, A. and Fontanot, F. and Franco, A. and García-Bellido, J. and Gasparetto, T. and Gavazzi, R. and Gaztanaga, E. and Giacomini, F. and Gianotti, F. and Gonzalez, A. H. and Gozaliasl, G. and Gruppuso, A. and Guidi, M. and Gutierrez, C. M. and Hall, A. and Hernández-Monteagudo, C. and Hildebrandt, H. and Hjorth, J. and Jacobson, J. and Joudaki, S. and Kajava, J. J. E. and Kang, Y. and Kansal, V. and Karagiannis, D. and Kiiveri, K. and Kirkpatrick, C. C. and Kruk, S. and Lacasa, F. and Laigle, C. and Lattanzi, M. and Brun, V. Le and Graet, J. Le and Legrand, L. and Lembo, M. and Lepori, F. and Leroy, G. and Lesci, G. F. and Lesgourgues, J. and Leuzzi, L. and Liaudat, T. I. and Loureiro, A. and Magliocchetti, M. and Magnier, E. A. and Mancini, C. and Mannucci, F. and Maoli, R. and Martins, C. J. A. P. and Maurin, L. and McPartland, C. J. R. and Melin, J.-B. and Migliaccio, M. and Miluzio, M. and Monaco, P. and Montoro, A. and Moretti, C. and Morgante, G. and Murray, C. and Nadathur, S. and Naidoo, K. and Navarro-Alsina, A. and Nesseris, S. and Nicastro, L. and Oguri, M. and Passalacqua, F. and Paterson, K. and Patrizii, L. and Pisani, A. and Potter, D. and Quai, S. and Radovich, M. and Reimberg, P. and Rocci, P.-F. and Rodighiero, G. and Rollins, R. P. and Sacquegna, S. and Sahlén, M. and Sanders, D. B. and Sarpa, E. and Scarlata, C. and Schaye, J. and Schneider, A. and Schultheis, M. and Sciotti, D. and Scognamiglio, D. and Sellentin, E. and Shankar, F. and Smith, L. C. and Soubrie, E. and Stanford, S. A. and Tanidis, K. and Tao, C. and Testera, G. and Tewes, M. and Teyssier, R. and Tosi, S. and Troja, A. and Tucci, M. and Valieri, C. and Venhola, A. and Vergani, D. and Vernizzi, F. and Verza, G. and Vielzeuf, P. and Walton, N. A. and Weaver, J. R. and Wilde, J. and Zalesky, L.},
	month = mar,
	year = {2025},
	note = {arXiv:2503.15302 [astro-ph]},
}

@misc{euclid_collaboration_euclid_2025-1,
	title = {Euclid {Quick} {Data} {Release} ({Q1}) -- {Characteristics} and limitations of the spectroscopic measurements},
	url = {http://arxiv.org/abs/2503.15308},
	doi = {10.48550/arXiv.2503.15308},
	urldate = {2025-07-31},
	publisher = {arXiv},
	author = {{Euclid Collaboration} and Brun, V. Le and Bethermin, M. and Moresco, M. and Vibert, D. and Vergani, D. and Surace, C. and Zamorani, G. and Allaoui, A. and Bedrine, T. and Chabaud, P.-Y. and Daste, G. and Dufresne, F. and Gray, M. and Rossetti, E. and Copin, Y. and Conseil, S. and Maiorano, E. and Mao, Z. and Palazzi, E. and Pozzetti, L. and Quai, S. and Scarlata, C. and Talia, M. and Courtois, H. M. and Guzzo, L. and Kubik, B. and Brun, A. M. C. Le and Peacock, J. A. and Scott, D. and Bagot, D. and Basset, A. and Casenove, P. and Gimenez, R. and Libet, G. and Ruffenach, M. and Aghanim, N. and Altieri, B. and Amara, A. and Andreon, S. and Auricchio, N. and Aussel, H. and Baccigalupi, C. and Baldi, M. and Balestra, A. and Bardelli, S. and Battaglia, P. and Biviano, A. and Bonchi, A. and Bonino, D. and Branchini, E. and Brescia, M. and Brinchmann, J. and Caillat, A. and Camera, S. and Cañas-Herrera, G. and Capobianco, V. and Carbone, C. and Carretero, J. and Casas, S. and Castander, F. J. and Castignani, G. and Cavuoti, S. and Chambers, K. C. and Cimatti, A. and Colodro-Conde, C. and Congedo, G. and Conselice, C. J. and Conversi, L. and Costille, A. and Courbin, F. and Cuby, J.-G. and Silva, A. Da and Degaudenzi, H. and Torre, S. de la and Lucia, G. De and Giorgio, A. M. Di and Dole, H. and Douspis, M. and Dubath, F. and Dupac, X. and Dusini, S. and Ealet, A. and Escoffier, S. and Fabricius, M. and Farina, M. and Farinelli, R. and Faustini, F. and Ferriol, S. and Fotopoulou, S. and Fourmanoit, N. and Frailis, M. and Franceschi, E. and Fumana, M. and Galeotta, S. and George, K. and Gillard, W. and Gillis, B. and Giocoli, C. and Gracia-Carpio, J. and Granett, B. R. and Grazian, A. and Grupp, F. and Haugan, S. V. H. and Hoar, J. and Hoekstra, H. and Holmes, W. and Hormuth, F. and Hornstrup, A. and Hudelot, P. and Jahnke, K. and Jhabvala, M. and Joachimi, B. and Keihänen, E. and Kermiche, S. and Kiessling, A. and Kümmel, M. and Kunz, M. and Kurki-Suonio, H. and Boulc'h, Q. Le and Mignant, D. Le and Ligori, S. and Lilje, P. B. and Lindholm, V. and Lloro, I. and Mainetti, G. and Maino, D. and Mansutti, O. and Marcin, S. and Marggraf, O. and Martinelli, M. and Martinet, N. and Marulli, F. and Massey, R. and Maurogordato, S. and Medinaceli, E. and Mei, S. and Melchior, M. and Mellier, Y. and Meneghetti, M. and Merlin, E. and Meylan, G. and Mora, A. and Moscardini, L. and Nakajima, R. and Neissner, C. and Nichol, R. C. and Niemi, S.-M. and Nightingale, J. W. and Padilla, C. and Paltani, S. and Pasian, F. and Pedersen, K. and Percival, W. J. and Pettorino, V. and Pires, S. and Polenta, G. and Poncet, M. and Popa, L. A. and Raison, F. and Rebolo, R. and Renzi, A. and Rhodes, J. and Riccio, G. and Romelli, E. and Roncarelli, M. and Saglia, R. and Sakr, Z. and Sapone, D. and Sartoris, B. and Sauvage, M. and Schewtschenko, J. A. and Schirmer, M. and Schneider, P. and Schrabback, T. and Scodeggio, M. and Secroun, A. and Seidel, G. and Seiffert, M. and Sirignano, C. and Sirri, G. and Stanco, L. and Steinwagner, J. and Tallada-Crespí, P. and Taylor, A. N. and Teplitz, H. I. and Tereno, I. and Tessore, N. and Toft, S. and Toledo-Moreo, R. and Torradeflot, F. and Tutusaus, I. and Valenziano, L. and Valiviita, J. and Vassallo, T. and Kleijn, G. Verdoes and Veropalumbo, A. and Wang, Y. and Weller, J. and Zacchei, A. and Zerbi, F. M. and Zinchenko, I. A. and Zucca, E. and Allevato, V. and Ballardini, M. and Bolzonella, M. and Bozzo, E. and Burigana, C. and Cabanac, R. and Cappi, A. and Ferdinando, D. Di and Vigo, J. A. Escartin and Fabbian, G. and Gabarra, L. and Hartley, W. G. and Martín-Fleitas, J. and Matthew, S. and Maturi, M. and Mauri, N. and Metcalf, R. B. and Pezzotta, A. and Pöntinen, M. and Porciani, C. and Risso, I. and Scottez, V. and Sereno, M. and Tenti, M. and Viel, M. and Wiesmann, M. and Akrami, Y. and Alvi, S. and Andika, I. T. and Anselmi, S. and Archidiacono, M. and Atrio-Barandela, F. and Avila, S. and Bella, M. and Bergamini, P. and Bertacca, D. and Blot, L. and Borgani, S. and Brown, M. L. and Bruton, S. and Calabro, A. and Quevedo, B. Camacho and Caro, F. and Carvalho, C. S. and Castro, T. and Charles, Y. and Chary, R. and Cogato, F. and Cooray, A. R. and Cucciati, O. and Davini, S. and Paolis, F. De and Desprez, G. and Díaz-Sánchez, A. and Diaz, J. J. and Domizio, S. Di and Diego, J. M. and Dimauro, P. and Duc, P.-A. and Enia, A. and Fang, Y. and Ferguson, A. M. N. and Ferrari, A. G. and Finoguenov, A. and Fontana, A. and Franco, A. and Ganga, K. and García-Bellido, J. and Gasparetto, T. and Gautard, V. and Gaztanaga, E. and Giacomini, F. and Gianotti, F. and Gozaliasl, G. and Gregorio, A. and Guidi, M. and Gutierrez, C. M. and Hall, A. and Hernández-Monteagudo, C. and Hildebrandt, H. and Hjorth, J. and Kajava, J. J. E. and Kang, Y. and Kansal, V. and Karagiannis, D. and Kiiveri, K. and Kirkpatrick, C. C. and Kruk, S. and Legrand, L. and Lembo, M. and Lepori, F. and Lesci, G. F. and Lesgourgues, J. and Leuzzi, L. and Liaudat, T. I. and Liu, S. J. and Loureiro, A. and Macias-Perez, J. and Magliocchetti, M. and Magnier, E. A. and Mancini, C. and Mannucci, F. and Maoli, R. and Martins, C. J. A. P. and Maurin, L. and Miluzio, M. and Monaco, P. and Montoro, A. and Moretti, C. and Morgante, G. and Nadathur, S. and Naidoo, K. and Navarro-Alsina, A. and Nesseris, S. and Passalacqua, F. and Paterson, K. and Patrizii, L. and Pisani, A. and Potter, D. and Radovich, M. and Rocci, P.-F. and Sacquegna, S. and Sahlén, M. and Sanders, D. B. and Sarpa, E. and Schneider, A. and Sciotti, D. and Sellentin, E. and Shankar, F. and Smith, L. C. and Tanidis, K. and Testera, G. and Teyssier, R. and Tosi, S. and Troja, A. and Tucci, M. and Valieri, C. and Venhola, A. and Verza, G. and Vielzeuf, P. and Walton, N. A. and Weaver, J. R. and Zalesky, L. and Sorce, J. G.},
	month = mar,
	year = {2025},
	note = {arXiv:2503.15308 [astro-ph]},
}

@article{euclid_collaboration_euclid_2025-2,
	title = {Euclid. {I}. {Overview} of the {Euclid} mission},
	volume = {697},
	issn = {0004-6361, 1432-0746},
	url = {http://arxiv.org/abs/2405.13491},
	doi = {10.1051/0004-6361/202450810},
	urldate = {2025-07-31},
	journal = {Astronomy \& Astrophysics},
	author = {{Euclid Collaboration} and Mellier, Y. and Abdurro'uf and Barroso, J. A. Acevedo and Achúcarro, A. and Adamek, J. and Adam, R. and Addison, G. E. and Aghanim, N. and Aguena, M. and Ajani, V. and Akrami, Y. and Al-Bahlawan, A. and Alavi, A. and Albuquerque, I. S. and Alestas, G. and Alguero, G. and Allaoui, A. and Allen, S. W. and Allevato, V. and Alonso-Tetilla, A. V. and Altieri, B. and Alvarez-Candal, A. and Alvi, S. and Amara, A. and Amendola, L. and Amiaux, J. and Andika, I. T. and Andreon, S. and Andrews, A. and Angora, G. and Angulo, R. E. and Annibali, F. and Anselmi, A. and Anselmi, S. and Arcari, S. and Archidiacono, M. and Aricò, G. and Arnaud, M. and Arnouts, S. and Asgari, M. and Asorey, J. and Atayde, L. and Atek, H. and Atrio-Barandela, F. and Aubert, M. and Aubourg, E. and Auphan, T. and Auricchio, N. and Aussel, B. and Aussel, H. and Avelino, P. P. and Avgoustidis, A. and Avila, S. and Awan, S. and Azzollini, R. and Baccigalupi, C. and Bachelet, E. and Bacon, D. and Baes, M. and Bagley, M. B. and Bahr-Kalus, B. and Balaguera-Antolinez, A. and Balbinot, E. and Balcells, M. and Baldi, M. and Baldry, I. and Balestra, A. and Ballardini, M. and Ballester, O. and Balogh, M. and Bañados, E. and Barbier, R. and Bardelli, S. and Baron, M. and Barreiro, T. and Barrena, R. and Barriere, J.-C. and Barros, B. J. and Barthelemy, A. and Bartolo, N. and Basset, A. and Battaglia, P. and Battisti, A. J. and Baugh, C. M. and Baumont, L. and Bazzanini, L. and Beaulieu, J.-P. and Beckmann, V. and Belikov, A. N. and Bel, J. and Bellagamba, F. and Bella, M. and Bellini, E. and Benabed, K. and Bender, R. and Benevento, G. and Bennett, C. L. and Benson, K. and Bergamini, P. and Bermejo-Climent, J. R. and Bernardeau, F. and Bertacca, D. and Berthe, M. and Berthier, J. and Bethermin, M. and Beutler, F. and Bevillon, C. and Bhargava, S. and Bhatawdekar, R. and Bianchi, D. and Bisigello, L. and Biviano, A. and Blake, R. P. and Blanchard, A. and Blazek, J. and Blot, L. and Bosco, A. and Bodendorf, C. and Boenke, T. and Böhringer, H. and Boldrini, P. and Bolzonella, M. and Bonchi, A. and Bonici, M. and Bonino, D. and Bonino, L. and Bonvin, C. and Bon, W. and Booth, J. T. and Borgani, S. and Borlaff, A. S. and Borsato, E. and Bosco, A. and Bose, B. and Botticella, M. T. and Boucaud, A. and Bouche, F. and Boucher, J. S. and Boutigny, D. and Bouvard, T. and Bouwens, R. and Bouy, H. and Bowler, R. A. A. and Bozza, V. and Bozzo, E. and Branchini, E. and Brando, G. and Brau-Nogue, S. and Brekke, P. and Bremer, M. N. and Brescia, M. and Breton, M.-A. and Brinchmann, J. and Brinckmann, T. and Brockley-Blatt, C. and Brodwin, M. and Brouard, L. and Brown, M. L. and Bruton, S. and Bucko, J. and Buddelmeijer, H. and Buenadicha, G. and Buitrago, F. and Burger, P. and Burigana, C. and Busillo, V. and Busonero, D. and Cabanac, R. and Cabayol-Garcia, L. and Cagliari, M. S. and Caillat, A. and Caillat, L. and Calabrese, M. and Calabro, A. and Calderone, G. and Calura, F. and Quevedo, B. Camacho and Camera, S. and Campos, L. and Canas-Herrera, G. and Candini, G. P. and Cantiello, M. and Capobianco, V. and Cappellaro, E. and Cappelluti, N. and Cappi, A. and Caputi, K. I. and Cara, C. and Carbone, C. and Cardone, V. F. and Carella, E. and Carlberg, R. G. and Carle, M. and Carminati, L. and Caro, F. and Carrasco, J. M. and Carretero, J. and Carrilho, P. and Duque, J. Carron and Carry, B. and Carvalho, A. and Carvalho, C. S. and Casas, R. and Casas, S. and Casenove, P. and Casey, C. M. and Cassata, P. and Castander, F. J. and Castelao, D. and Castellano, M. and Castiblanco, L. and Castignani, G. and Castro, T. and Cavet, C. and Cavuoti, S. and Chabaud, P.-Y. and Chambers, K. C. and Charles, Y. and Charlot, S. and Chartab, N. and Chary, R. and Chaumeil, F. and Cho, H. and Chon, G. and Ciancetta, E. and Ciliegi, P. and Cimatti, A. and Cimino, M. and Cioni, M.-R. L. and Claydon, R. and Cleland, C. and Clément, B. and Clements, D. L. and Clerc, N. and Clesse, S. and Codis, S. and Cogato, F. and Colbert, J. and Cole, R. E. and Coles, P. and Collett, T. E. and Collins, R. S. and Colodro-Conde, C. and Colombo, C. and Combes, F. and Conforti, V. and Congedo, G. and Conseil, S. and Conselice, C. J. and Contarini, S. and Contini, T. and Conversi, L. and Cooray, A. R. and Copin, Y. and Corasaniti, P.-S. and Corcho-Caballero, P. and Corcione, L. and Cordes, O. and Corpace, O. and Correnti, M. and Costanzi, M. and Costille, A. and Courbin, F. and Mifsud, L. Courcoult and Courtois, H. M. and Cousinou, M.-C. and Covone, G. and Cowell, T. and Cragg, C. and Cresci, G. and Cristiani, S. and Crocce, M. and Cropper, M. and Crouzet, P. E. and Csizi, B. and Cuby, J.-G. and Cucchetti, E. and Cucciati, O. and Cuillandre, J.-C. and Cunha, P. A. C. and Cuozzo, V. and Daddi, E. and D'Addona, M. and Dafonte, C. and Dagoneau, N. and Dalessandro, E. and Dalton, G. B. and D'Amico, G. and Dannerbauer, H. and Danto, P. and Das, I. and Silva, A. Da and Silva, R. da and Doumerg, W. d'Assignies and Daste, G. and Davies, J. E. and Davini, S. and Dayal, P. and Boer, T. de and Decarli, R. and Caro, B. De and Degaudenzi, H. and Degni, G. and Jong, J. T. A. de and Bella, L. F. de la and Torre, S. de la and Delhaise, F. and Delley, D. and Delucchi, G. and Lucia, G. De and Denniston, J. and Paolis, F. De and Petris, M. De and Derosa, A. and Desai, S. and Desjacques, V. and Despali, G. and Desprez, G. and Vicente-Albendea, J. De and Deville, Y. and Dias, J. D. F. and Díaz-Sánchez, A. and Diaz, J. J. and Domizio, S. Di and Diego, J. M. and Ferdinando, D. Di and Giorgio, A. M. Di and Dimauro, P. and Dinis, J. and Dolag, K. and Dolding, C. and Dole, H. and Sánchez, H. Domínguez and Doré, O. and Dournac, F. and Douspis, M. and Dreihahn, H. and Droge, B. and Dryer, B. and Dubath, F. and Duc, P.-A. and Ducret, F. and Duffy, C. and Dufresne, F. and Duncan, C. A. J. and Dupac, X. and Duret, V. and Durrer, R. and Durret, F. and Dusini, S. and Ealet, A. and Eggemeier, A. and Eisenhardt, P. R. M. and Elbaz, D. and Elkhashab, M. Y. and Ellien, A. and Endicott, J. and Enia, A. and Erben, T. and Vigo, J. A. Escartin and Escoffier, S. and Sanz, I. Escudero and Essert, J. and Ettori, S. and Ezziati, M. and Fabbian, G. and Fabricius, M. and Fang, Y. and Farina, A. and Farina, M. and Farinelli, R. and Farrens, S. and Faustini, F. and Feltre, A. and Ferguson, A. M. N. and Ferrando, P. and Ferrari, A. G. and Ferré-Mateu, A. and Ferreira, P. G. and Ferreras, I. and Ferrero, I. and Ferriol, S. and Ferruit, P. and Filleul, D. and Finelli, F. and Finkelstein, S. L. and Finoguenov, A. and Fiorini, B. and Flentge, F. and Focardi, P. and Fonseca, J. and Fontana, A. and Fontanot, F. and Fornari, F. and Fosalba, P. and Fossati, M. and Fotopoulou, S. and Fouchez, D. and Fourmanoit, N. and Frailis, M. and Fraix-Burnet, D. and Franceschi, E. and Franco, A. and Franzetti, P. and Freihoefer, J. and Frenk, C. S. and Frittoli, G. and Frugier, P.-A. and Frusciante, N. and Fumagalli, A. and Fumagalli, M. and Fumana, M. and Fu, Y. and Gabarra, L. and Galeotta, S. and Galluccio, L. and Ganga, K. and Gao, H. and García-Bellido, J. and Garcia, K. and Gardner, J. P. and Garilli, B. and Gaspar-Venancio, L.-M. and Gasparetto, T. and Gautard, V. and Gavazzi, R. and Gaztanaga, E. and Genolet, L. and Santos, R. Genova and Gentile, F. and George, K. and Gerbino, M. and Ghaffari, Z. and Giacomini, F. and Gianotti, F. and Gibb, G. P. S. and Gillard, W. and Gillis, B. and Ginolfi, M. and Giocoli, C. and Girardi, M. and Giri, S. K. and Goh, L. W. K. and Gómez-Alvarez, P. and Gonzalez-Perez, V. and Gonzalez, A. H. and Gonzalez, E. J. and Gonzalez, J. C. and Beauchamps, S. Gouyou and Gozaliasl, G. and Gracia-Carpio, J. and Grandis, S. and Granett, B. R. and Granvik, M. and Grazian, A. and Gregorio, A. and Grenet, C. and Grillo, C. and Grupp, F. and Gruppioni, C. and Gruppuso, A. and Guerbuez, C. and Guerrini, S. and Guidi, M. and Guillard, P. and Gutierrez, C. M. and Guttridge, P. and Guzzo, L. and Gwyn, S. and Haapala, J. and Haase, J. and Haddow, C. R. and Hailey, M. and Hall, A. and Hall, D. and Hamaus, N. and Haridasu, B. S. and Harnois-Déraps, J. and Harper, C. and Hartley, W. G. and Hasinger, G. and Hassani, F. and Hatch, N. A. and Haugan, S. V. H. and Häußler, B. and Heavens, A. and Heisenberg, L. and Helmi, A. and Helou, G. and Hemmati, S. and Henares, K. and Herent, O. and Hernández-Monteagudo, C. and Heuberger, T. and Hewett, P. C. and Heydenreich, S. and Hildebrandt, H. and Hirschmann, M. and Hjorth, J. and Hoar, J. and Hoekstra, H. and Holland, A. D. and Holliman, M. S. and Holmes, W. and Hook, I. and Horeau, B. and Hormuth, F. and Hornstrup, A. and Hosseini, S. and Hu, D. and Hudelot, P. and Hudson, M. J. and Huertas-Company, M. and Huff, E. M. and Hughes, A. C. N. and Humphrey, A. and Hunt, L. K. and Huynh, D. D. and Ibata, R. and Ichikawa, K. and Iglesias-Groth, S. and Ilbert, O. and Ilić, S. and Ingoglia, L. and Iodice, E. and Israel, H. and Israelsson, U. E. and Izzo, L. and Jablonka, P. and Jackson, N. and Jacobson, J. and Jafariyazani, M. and Jahnke, K. and Jain, B. and Jansen, H. and Jarvis, M. J. and Jasche, J. and Jauzac, M. and Jeffrey, N. and Jhabvala, M. and Jimenez-Teja, Y. and Muñoz, A. Jimenez and Joachimi, B. and Johansson, P. H. and Joudaki, S. and Jullo, E. and Kajava, J. J. E. and Kang, Y. and Kannawadi, A. and Kansal, V. and Karagiannis, D. and Kärcher, M. and Kashlinsky, A. and Kazandjian, M. V. and Keck, F. and Keihänen, E. and Kerins, E. and Kermiche, S. and Khalil, A. and Kiessling, A. and Kiiveri, K. and Kilbinger, M. and Kim, J. and King, R. and Kirkpatrick, C. C. and Kitching, T. and Kluge, M. and Knabenhans, M. and Knapen, J. H. and Knebe, A. and Kneib, J.-P. and Kohley, R. and Koopmans, L. V. E. and Koskinen, H. and Koulouridis, E. and Kou, R. and Kovács, A. and Kovačić, I. and Kowalczyk, A. and Koyama, K. and Kraljic, K. and Krause, O. and Kruk, S. and Kubik, B. and Kuchner, U. and Kuijken, K. and Kümmel, M. and Kunz, M. and Kurki-Suonio, H. and Lacasa, F. and Lacey, C. G. and Franca, F. La and Lagarde, N. and Lahav, O. and Laigle, C. and Marca, A. La and Marle, O. La and Lamine, B. and Lam, M. C. and Lançon, A. and Landt, H. and Langer, M. and Lapi, A. and Larcheveque, C. and Larsen, S. S. and Lattanzi, M. and Laudisio, F. and Laugier, D. and Laureijs, R. and Laurent, V. and Lavaux, G. and Lawrenson, A. and Lazanu, A. and Lazeyras, T. and Boulc'h, Q. Le and Brun, A. M. C. Le and Brun, V. Le and Leclercq, F. and Lee, S. and Graet, J. Le and Legrand, L. and Leirvik, K. N. and Jeune, M. Le and Lembo, M. and Mignant, D. Le and Lepinzan, M. D. and Lepori, F. and Reun, A. Le and Leroy, G. and Lesci, G. F. and Lesgourgues, J. and Leuzzi, L. and Levi, M. E. and Liaudat, T. I. and Libet, G. and Liebing, P. and Ligori, S. and Lilje, P. B. and Lin, C.-C. and Linde, D. and Linder, E. and Lindholm, V. and Linke, L. and Li, S.-S. and Liu, S. J. and Lloro, I. and Lobo, F. S. N. and Lodieu, N. and Lombardi, M. and Lombriser, L. and Lonare, P. and Longo, G. and López-Caniego, M. and Lopez, X. Lopez and Alvarez, J. Lorenzo and Loureiro, A. and Loveday, J. and Lusso, E. and Macias-Perez, J. and Maciaszek, T. and Maggio, G. and Magliocchetti, M. and Magnard, F. and Magnier, E. A. and Magro, A. and Mahler, G. and Mainetti, G. and Maino, D. and Maiorano, E. and Maiorano, E. and Malavasi, N. and Mamon, G. A. and Mancini, C. and Mandelbaum, R. and Manera, M. and Manjón-García, A. and Mannucci, F. and Mansutti, O. and Outeiro, M. Manteiga and Maoli, R. and Maraston, C. and Marcin, S. and Marcos-Arenal, P. and Margalef-Bentabol, B. and Marggraf, O. and Marinucci, D. and Marinucci, M. and Markovic, K. and Marleau, F. R. and Marpaud, J. and Martignac, J. and Martín-Fleitas, J. and Martin-Moruno, P. and Martin, E. L. and Martinelli, M. and Martinet, N. and Martin, H. and Martins, C. J. A. P. and Marulli, F. and Massari, D. and Massey, R. and Masters, D. C. and Matarrese, S. and Matsuoka, Y. and Matthew, S. and Maughan, B. J. and Mauri, N. and Maurin, L. and Maurogordato, S. and McCarthy, K. and McConnachie, A. W. and McCracken, H. J. and McDonald, I. and McEwen, J. D. and McPartland, C. J. R. and Medinaceli, E. and Mehta, V. and Mei, S. and Melchior, M. and Melin, J.-B. and Ménard, B. and Mendes, J. and Mendez-Abreu, J. and Meneghetti, M. and Mercurio, A. and Merlin, E. and Metcalf, R. B. and Meylan, G. and Migliaccio, M. and Mignoli, M. and Miller, L. and Miluzio, M. and Milvang-Jensen, B. and Mimoso, J. P. and Miquel, R. and Miyatake, H. and Mobasher, B. and Mohr, J. J. and Monaco, P. and Monguió, M. and Montoro, A. and Mora, A. and Dizgah, A. Moradinezhad and Moresco, M. and Moretti, C. and Morgante, G. and Morisset, N. and Moriya, T. J. and Morris, P. W. and Mortlock, D. J. and Moscardini, L. and Mota, D. F. and Mottet, S. and Moustakas, L. A. and Moutard, T. and Müller, T. and Munari, E. and Murphree, G. and Murray, C. and Murray, N. and Musi, P. and Nadathur, S. and Nagam, B. C. and Nagao, T. and Naidoo, K. and Nakajima, R. and Nally, C. and Natoli, P. and Navarro-Alsina, A. and Girones, D. Navarro and Neissner, C. and Nersesian, A. and Nesseris, S. and Nguyen-Kim, H. N. and Nicastro, L. and Nichol, R. C. and Nielbock, M. and Niemi, S.-M. and Nieto, S. and Nilsson, K. and Noller, J. and Norberg, P. and Nouri-Zonoz, A. and Ntelis, P. and Nucita, A. A. and Nugent, P. and Nunes, N. J. and Nutma, T. and Ocampo, I. and Odier, J. and Oesch, P. A. and Oguri, M. and Oliveira, D. Magalhaes and Onoue, M. and Oosterbroek, T. and Oppizzi, F. and Ordenovic, C. and Osato, K. and Pacaud, F. and Pace, F. and Padilla, C. and Paech, K. and Pagano, L. and Page, M. J. and Palazzi, E. and Paltani, S. and Pamuk, S. and Pandolfi, S. and Paoletti, D. and Paolillo, M. and Papaderos, P. and Pardede, K. and Parimbelli, G. and Parmar, A. and Partmann, C. and Pasian, F. and Passalacqua, F. and Paterson, K. and Patrizii, L. and Pattison, C. and Paulino-Afonso, A. and Paviot, R. and Peacock, J. A. and Pearce, F. R. and Pedersen, K. and Peel, A. and Peletier, R. F. and Ibanez, M. Pellejero and Pello, R. and Penny, M. T. and Percival, W. J. and Perez-Garrido, A. and Perotto, L. and Pettorino, V. and Pezzotta, A. and Pezzuto, S. and Philippon, A. and Pierre, M. and Piersanti, O. and Pietroni, M. and Piga, L. and Pilo, L. and Pires, S. and Pisani, A. and Pizzella, A. and Pizzuti, L. and Plana, C. and Polenta, G. and Pollack, J. E. and Poncet, M. and Pöntinen, M. and Pool, P. and Popa, L. A. and Popa, V. and Popp, J. and Porciani, C. and Porth, L. and Potter, D. and Poulain, M. and Pourtsidou, A. and Pozzetti, L. and Prandoni, I. and Pratt, G. W. and Prezelus, S. and Prieto, E. and Pugno, A. and Quai, S. and Quilley, L. and Racca, G. D. and Raccanelli, A. and Rácz, G. and Radinović, S. and Radovich, M. and Ragagnin, A. and Ragnit, U. and Raison, F. and Ramos-Chernenko, N. and Ranc, C. and Rasera, Y. and Raylet, N. and Rebolo, R. and Refregier, A. and Reimberg, P. and Reiprich, T. H. and Renk, F. and Renzi, A. and Retre, J. and Revaz, Y. and Reylé, C. and Reynolds, L. and Rhodes, J. and Ricci, F. and Ricci, M. and Riccio, G. and Ricken, S. O. and Rissanen, S. and Risso, I. and Rix, H.-W. and Robin, A. C. and Rocca-Volmerange, B. and Rocci, P.-F. and Rodenhuis, M. and Rodighiero, G. and Monroy, M. Rodriguez and Rollins, R. P. and Romanello, M. and Roman, J. and Romelli, E. and Romero-Gomez, M. and Roncarelli, M. and Rosati, P. and Rosset, C. and Rossetti, E. and Roster, W. and Rottgering, H. J. A. and Rozas-Fernández, A. and Ruane, K. and Rubino-Martin, J. A. and Rudolph, A. and Ruppin, F. and Rusholme, B. and Sacquegna, S. and Sáez-Casares, I. and Saga, S. and Saglia, R. and Sahlén, M. and Saifollahi, T. and Sakr, Z. and Salvalaggio, J. and Salvaterra, R. and Salvati, L. and Salvato, M. and Salvignol, J.-C. and Sánchez, A. G. and Sanchez, E. and Sanders, D. B. and Sapone, D. and Saponara, M. and Sarpa, E. and Sarron, F. and Sartori, S. and Sartoris, B. and Sassolas, B. and Sauniere, L. and Sauvage, M. and Sawicki, M. and Scaramella, R. and Scarlata, C. and Scharré, L. and Schaye, J. and Schewtschenko, J. A. and Schindler, J.-T. and Schinnerer, E. and Schirmer, M. and Schmidt, F. and Schmidt, F. and Schmidt, M. and Schneider, A. and Schneider, M. and Schneider, P. and Schöneberg, N. and Schrabback, T. and Schultheis, M. and Schulz, S. and Schuster, N. and Schwartz, J. and Sciotti, D. and Scodeggio, M. and Scognamiglio, D. and Scott, D. and Scottez, V. and Secroun, A. and Sefusatti, E. and Seidel, G. and Seiffert, M. and Sellentin, E. and Selwood, M. and Semboloni, E. and Sereno, M. and Serjeant, S. and Serrano, S. and Setnikar, G. and Shankar, F. and Sharples, R. M. and Short, A. and Shulevski, A. and Shuntov, M. and Sias, M. and Sikkema, G. and Silvestri, A. and Simon, P. and Sirignano, C. and Sirri, G. and Skottfelt, J. and Slezak, E. and Sluse, D. and Smith, G. P. and Smith, L. C. and Smith, R. E. and Smit, S. J. A. and Soldano, F. and Solheim, B. G. B. and Sorce, J. G. and Sorrenti, F. and Soubrie, E. and Spinoglio, L. and Mancini, A. Spurio and Stadel, J. and Stagnaro, L. and Stanco, L. and Stanford, S. A. and Starck, J.-L. and Stassi, P. and Steinwagner, J. and Stern, D. and Stone, C. and Strada, P. and Strafella, F. and Stramaccioni, D. and Surace, C. and Sureau, F. and Suyu, S. H. and Swindells, I. and Szafraniec, M. and Szapudi, I. and Taamoli, S. and Talia, M. and Tallada-Crespí, P. and Tanidis, K. and Tao, C. and Tarrío, P. and Tavagnacco, D. and Taylor, A. N. and Taylor, J. E. and Taylor, P. L. and Teixeira, E. M. and Tenti, M. and Idiago, P. Teodoro and Teplitz, H. I. and Tereno, I. and Tessore, N. and Testa, V. and Testera, G. and Tewes, M. and Teyssier, R. and Theret, N. and Thizy, C. and Thomas, P. D. and Toba, Y. and Toft, S. and Toledo-Moreo, R. and Tolstoy, E. and Tommasi, E. and Torbaniuk, O. and Torradeflot, F. and Tortora, C. and Tosi, S. and Tosti, S. and Trifoglio, M. and Troja, A. and Trombetti, T. and Tronconi, A. and Tsedrik, M. and Tsyganov, A. and Tucci, M. and Tutusaus, I. and Uhlemann, C. and Ulivi, L. and Urbano, M. and Vacher, L. and Vaillon, L. and Valageas, P. and Valdes, I. and Valentijn, E. A. and Valenziano, L. and Valieri, C. and Valiviita, J. and Broeck, M. Van den and Vassallo, T. and Vavrek, R. and Vega-Ferrero, J. and Venemans, B. and Venhola, A. and Ventura, S. and Kleijn, G. Verdoes and Vergani, D. and Verma, A. and Vernizzi, F. and Veropalumbo, A. and Verza, G. and Vescovi, C. and Vibert, D. and Viel, M. and Vielzeuf, P. and Viglione, C. and Viitanen, A. and Villaescusa-Navarro, F. and Vinciguerra, S. and Visticot, F. and Voggel, K. and Wietersheim-Kramsta, M. von and Vriend, W. J. and Wachter, S. and Walmsley, M. and Walth, G. and Walton, D. M. and Walton, N. A. and Wander, M. and Wang, L. and Wang, Y. and Weaver, J. R. and Weller, J. and Wetzstein, M. and Whalen, D. J. and Whittam, I. H. and Widmer, A. and Wiesmann, M. and Wilde, J. and Williams, O. R. and Winther, H.-A. and Wittje, A. and Wong, J. H. W. and Wright, A. H. and Yankelevich, V. and Yeung, H. W. and Yoon, M. and Youles, S. and Yung, L. Y. A. and Zacchei, A. and Zalesky, L. and Zamorani, G. and Vitorelli, A. Zamorano and Marc, M. Zanoni and Zennaro, M. and Zerbi, F. M. and Zinchenko, I. A. and Zoubian, J. and Zucca, E. and Zumalacarregui, M.},
	month = may,
	year = {2025},
	note = {arXiv:2405.13491 [astro-ph]},
	pages = {A1},
}

@misc{euclid_collaboration_euclid_2025-3,
	title = {Euclid {Quick} {Data} {Release} ({Q1}): {From} spectrograms to spectra: the {SIR} spectroscopic {Processing} {Function}},
	shorttitle = {Euclid {Quick} {Data} {Release} ({Q1})},
	url = {http://arxiv.org/abs/2503.15307},
	doi = {10.48550/arXiv.2503.15307},
	urldate = {2025-07-31},
	publisher = {arXiv},
	author = {{Euclid Collaboration} and Copin, Y. and Fumana, M. and Mancini, C. and Appleton, P. N. and Chary, R. and Conseil, S. and Faisst, A. L. and Hemmati, S. and Masters, D. C. and Scarlata, C. and Scodeggio, M. and Alavi, A. and Carle, A. and Casenove, P. and Contini, T. and Das, I. and Gillard, W. and Herzog, G. and Jacobson, J. and Brun, V. Le and Maino, D. and Setnikar, G. and Stickley, N. R. and Tavagnacco, D. and Xie, Q. and Aghanim, N. and Altieri, B. and Amara, A. and Andreon, S. and Auricchio, N. and Aussel, H. and Baccigalupi, C. and Baldi, M. and Balestra, A. and Bardelli, S. and Basset, A. and Battaglia, P. and Belikov, A. N. and Biviano, A. and Bonchi, A. and Branchini, E. and Brescia, M. and Brinchmann, J. and Camera, S. and Cañas-Herrera, G. and Capobianco, V. and Carbone, C. and Carretero, J. and Casas, S. and Castander, F. J. and Castellano, M. and Castignani, G. and Cavuoti, S. and Chambers, K. C. and Cimatti, A. and Colodro-Conde, C. and Congedo, G. and Conselice, C. J. and Conversi, L. and Courbin, F. and Courtois, H. M. and Silva, A. Da and Silva, R. da and Degaudenzi, H. and Torre, S. de la and Lucia, G. De and Giorgio, A. M. Di and Dole, H. and Dubath, F. and Dupac, X. and Dusini, S. and Ealet, A. and Escoffier, S. and Farina, M. and Farinelli, R. and Ferriol, S. and Finelli, F. and Fotopoulou, S. and Fourmanoit, N. and Frailis, M. and Franceschi, E. and Franzetti, P. and Galeotta, S. and George, K. and Gillis, B. and Giocoli, C. and Gracia-Carpio, J. and Granett, B. R. and Grazian, A. and Grupp, F. and Guzzo, L. and Haugan, S. V. H. and Hoar, J. and Hoekstra, H. and Holmes, W. and Hook, I. M. and Hormuth, F. and Hornstrup, A. and Hudelot, P. and Jahnke, K. and Jhabvala, M. and Joachimi, B. and Keihänen, E. and Kermiche, S. and Kiessling, A. and Kubik, B. and Kuijken, K. and Kümmel, M. and Kunz, M. and Kurki-Suonio, H. and Boulc'h, Q. Le and Brun, A. M. C. Le and Mignant, D. Le and Ligori, S. and Lilje, P. B. and Lindholm, V. and Lloro, I. and Mainetti, G. and Maiorano, E. and Mansutti, O. and Marcin, S. and Marggraf, O. and Markovic, K. and Martinelli, M. and Martinet, N. and Marulli, F. and Massey, R. and Maurogordato, S. and Medinaceli, E. and Mei, S. and Melchior, M. and Mellier, Y. and Meneghetti, M. and Merlin, E. and Meylan, G. and Mora, A. and Moresco, M. and Morris, P. W. and Moscardini, L. and Nakajima, R. and Neissner, C. and Nichol, R. C. and Niemi, S.-M. and Nightingale, J. W. and Padilla, C. and Paltani, S. and Pasian, F. and Pedersen, K. and Percival, W. J. and Pettorino, V. and Pires, S. and Polenta, G. and Poncet, M. and Popa, L. A. and Pozzetti, L. and Racca, G. D. and Raison, F. and Rebolo, R. and Renzi, A. and Rhodes, J. and Riccio, G. and Romelli, E. and Roncarelli, M. and Rossetti, E. and Saglia, R. and Sakr, Z. and Sánchez, A. G. and Sapone, D. and Sartoris, B. and Schewtschenko, J. A. and Schirmer, M. and Schneider, P. and Schrabback, T. and Secroun, A. and Sefusatti, E. and Seidel, G. and Serrano, S. and Simon, P. and Sirignano, C. and Sirri, G. and Mancini, A. Spurio and Stanco, L. and Steinwagner, J. and Tallada-Crespí, P. and Taylor, A. N. and Teplitz, H. I. and Tereno, I. and Tessore, N. and Toft, S. and Toledo-Moreo, R. and Torradeflot, F. and Tutusaus, I. and Valenziano, L. and Valiviita, J. and Vassallo, T. and Kleijn, G. Verdoes and Veropalumbo, A. and Wang, Y. and Weller, J. and Zacchei, A. and Zamorani, G. and Zerbi, F. M. and Zinchenko, I. A. and Zucca, E. and Allevato, V. and Ballardini, M. and Bolzonella, M. and Bozzo, E. and Burigana, C. and Cabanac, R. and Cappi, A. and Ferdinando, D. Di and Vigo, J. A. Escartin and Gabarra, L. and Huertas-Company, M. and Martín-Fleitas, J. and Matthew, S. and Mauri, N. and Metcalf, R. B. and Pezzotta, A. and Pöntinen, M. and Porciani, C. and Risso, I. and Scottez, V. and Sereno, M. and Tenti, M. and Viel, M. and Wiesmann, M. and Akrami, Y. and Alvi, S. and Andika, I. T. and Anselmi, S. and Archidiacono, M. and Atrio-Barandela, F. and Benoist, C. and Benson, K. and Bergamini, P. and Bertacca, D. and Bethermin, M. and Bisigello, L. and Blanchard, A. and Blot, L. and Brown, M. L. and Bruton, S. and Calabro, A. and Quevedo, B. Camacho and Caro, F. and Castro, T. and Cogato, F. and Cooray, A. R. and Cucciati, O. and Davini, S. and Paolis, F. De and Desprez, G. and Díaz-Sánchez, A. and Diaz, J. J. and Domizio, S. Di and Diego, J. M. and Duc, P.-A. and Enia, A. and Fang, Y. and Ferguson, A. M. N. and Ferrari, A. G. and Finoguenov, A. and Fontana, A. and Franco, A. and Ganga, K. and García-Bellido, J. and Gasparetto, T. and Gautard, V. and Gaztanaga, E. and Giacomini, F. and Gianotti, F. and Gozaliasl, G. and Gregorio, A. and Guidi, M. and Gutierrez, C. M. and Hall, A. and Hartley, W. G. and Hernández-Monteagudo, C. and Hildebrandt, H. and Hjorth, J. and Hosseini, S. and Kajava, J. J. E. and Kang, Y. and Kansal, V. and Karagiannis, D. and Kiiveri, K. and Kirkpatrick, C. C. and Kruk, S. and Graet, J. Le and Legrand, L. and Lembo, M. and Lepori, F. and Leroy, G. and Lesci, G. F. and Lesgourgues, J. and Leuzzi, L. and Liaudat, T. I. and Loureiro, A. and Macias-Perez, J. and Maggio, G. and Magliocchetti, M. and Mannucci, F. and Maoli, R. and Martins, C. J. A. P. and Maurin, L. and McPartland, C. J. R. and Miluzio, M. and Monaco, P. and Montoro, A. and Moretti, C. and Morgante, G. and Murray, C. and Nadathur, S. and Naidoo, K. and Navarro-Alsina, A. and Nesseris, S. and Passalacqua, F. and Paterson, K. and Patrizii, L. and Pisani, A. and Potter, D. and Quai, S. and Radovich, M. and Rocci, P.-F. and Rodighiero, G. and Sacquegna, S. and Sahlén, M. and Sanders, D. B. and Sarpa, E. and Schneider, A. and Sciotti, D. and Sellentin, E. and Smith, L. C. and Tanidis, K. and Tao, C. and Testera, G. and Teyssier, R. and Tosi, S. and Troja, A. and Tucci, M. and Valieri, C. and Venhola, A. and Vergani, D. and Verza, G. and Vielzeuf, P. and Walton, N. A. and Bella, M. and Scott, D.},
	month = mar,
	year = {2025},
	note = {arXiv:2503.15307 [astro-ph]},
}

@misc{euclid_collaboration_euclid_2025-5,
	title = {Euclid preparation. {The} impact of redshift interlopers on the two-point correlation function analysis},
	url = {http://arxiv.org/abs/2505.04688},
	doi = {10.48550/arXiv.2505.04688},
	urldate = {2025-07-31},
	publisher = {arXiv},
	author = {{Euclid Collaboration} and Risso, I. and Veropalumbo, A. and Branchini, E. and Maragliano, E. and Torre, S. de la and Sarpa, E. and Monaco, P. and Granett, B. R. and Lee, S. and Addison, G. E. and Bruton, S. and Carbone, C. and Lavaux, G. and Markovic, K. and McCarthy, K. and Parimbelli, G. and Passalacqua, F. and Percival, W. J. and Scarlata, C. and Sefusatti, E. and Wang, Y. and Bonici, M. and Oppizzi, F. and Aghanim, N. and Altieri, B. and Amara, A. and Andreon, S. and Auricchio, N. and Baccigalupi, C. and Baldi, M. and Balestra, A. and Bardelli, S. and Battaglia, P. and Biviano, A. and Bonchi, A. and Bonino, D. and Brescia, M. and Brinchmann, J. and Camera, S. and Cañas-Herrera, G. and Capobianco, V. and Cardone, V. F. and Carretero, J. and Casas, S. and Castellano, M. and Castignani, G. and Cavuoti, S. and Chambers, K. C. and Cimatti, A. and Colodro-Conde, C. and Congedo, G. and Conselice, C. J. and Conversi, L. and Copin, Y. and Courbin, F. and Courtois, H. M. and Crocce, M. and Silva, A. Da and Degaudenzi, H. and Lucia, G. De and Giorgio, A. M. Di and Dole, H. and Douspis, M. and Dubath, F. and Duncan, C. A. J. and Dupac, X. and Dusini, S. and Escoffier, S. and Farina, M. and Farinelli, R. and Faustini, F. and Ferriol, S. and Finelli, F. and Fotopoulou, S. and Fourmanoit, N. and Frailis, M. and Franceschi, E. and Fumana, M. and Galeotta, S. and George, K. and Gillard, W. and Gillis, B. and Giocoli, C. and Gracia-Carpio, J. and Grazian, A. and Grupp, F. and Guzzo, L. and Haugan, S. V. H. and Holmes, W. and Hormuth, F. and Hornstrup, A. and Hudelot, P. and Jahnke, K. and Jhabvala, M. and Joachimi, B. and Keihänen, E. and Kermiche, S. and Kiessling, A. and Kilbinger, M. and Kubik, B. and Kümmel, M. and Kunz, M. and Kurki-Suonio, H. and Brun, A. M. C. Le and Liebing, P. and Ligori, S. and Lilje, P. B. and Lindholm, V. and Lloro, I. and Mainetti, G. and Maino, D. and Maiorano, E. and Mansutti, O. and Marcin, S. and Marggraf, O. and Martinelli, M. and Martinet, N. and Marulli, F. and Massey, R. and Maurogordato, S. and Medinaceli, E. and Mei, S. and Melchior, M. and Mellier, Y. and Meneghetti, M. and Merlin, E. and Meylan, G. and Mora, A. and Moresco, M. and Moscardini, L. and Nakajima, R. and Neissner, C. and Niemi, S.-M. and Nightingale, J. W. and Padilla, C. and Paltani, S. and Pasian, F. and Pedersen, K. and Pettorino, V. and Pires, S. and Polenta, G. and Poncet, M. and Popa, L. A. and Pozzetti, L. and Raison, F. and Rebolo, R. and Renzi, A. and Rhodes, J. and Riccio, G. and Romelli, E. and Roncarelli, M. and Rossetti, E. and Saglia, R. and Sakr, Z. and Sapone, D. and Sartoris, B. and Schewtschenko, J. A. and Schneider, P. and Schrabback, T. and Scodeggio, M. and Secroun, A. and Seidel, G. and Seiffert, M. and Serrano, S. and Simon, P. and Sirignano, C. and Sirri, G. and Stanco, L. and Steinwagner, J. and Surace, C. and Tallada-Crespí, P. and Tavagnacco, D. and Taylor, A. N. and Tereno, I. and Tessore, N. and Toft, S. and Toledo-Moreo, R. and Torradeflot, F. and Tutusaus, I. and Valenziano, L. and Valiviita, J. and Vassallo, T. and Kleijn, G. Verdoes and Vibert, D. and Weller, J. and Zamorani, G. and Zerbi, F. M. and Zucca, E. and Allevato, V. and Ballardini, M. and Bolzonella, M. and Bozzo, E. and Burigana, C. and Cabanac, R. and Cappi, A. and Ferdinando, D. Di and Vigo, J. A. Escartin and Gabarra, L. and Hartley, W. G. and Martín-Fleitas, J. and Matthew, S. and Mauri, N. and Metcalf, R. B. and Pezzotta, A. and Pöntinen, M. and Porciani, C. and Scottez, V. and Sereno, M. and Tenti, M. and Viel, M. and Wiesmann, M. and Akrami, Y. and Alvi, S. and Andika, I. T. and Archidiacono, M. and Atrio-Barandela, F. and Avila, S. and Balaguera-Antolinez, A. and Benoist, C. and Bertacca, D. and Bethermin, M. and Blot, L. and Böhringer, H. and Borgani, S. and Brown, M. L. and Calabro, A. and Quevedo, B. Camacho and Caro, F. and Carvalho, C. S. and Castro, T. and Cogato, F. and Cooray, A. R. and Cucciati, O. and Davini, S. and Paolis, F. De and Desprez, G. and Díaz-Sánchez, A. and Diaz, J. J. and Domizio, S. Di and Diego, J. M. and Dimauro, P. and Enia, A. and Fang, Y. and Ferrari, A. G. and Finoguenov, A. and Fontana, A. and Franco, A. and Ganga, K. and García-Bellido, J. and Gasparetto, T. and Gautard, V. and Gaztanaga, E. and Giacomini, F. and Gianotti, F. and Gozaliasl, G. and Guidi, M. and Gutierrez, C. M. and Hall, A. and Hemmati, S. and Hernández-Monteagudo, C. and Hildebrandt, H. and Hjorth, J. and Joudaki, S. and Kajava, J. J. E. and Kang, Y. and Kansal, V. and Karagiannis, D. and Kiiveri, K. and Kirkpatrick, C. C. and Kruk, S. and Brun, V. Le and Graet, J. Le and Legrand, L. and Lembo, M. and Lepori, F. and Leroy, G. and Lesci, G. F. and Leuzzi, L. and Liaudat, T. I. and Loureiro, A. and Macias-Perez, J. and Magliocchetti, M. and Mannucci, F. and Maoli, R. and Martins, C. J. A. P. and Maurin, L. and Miluzio, M. and Moretti, C. and Morgante, G. and Nadathur, S. and Naidoo, K. and Navarro-Alsina, A. and Paterson, K. and Patrizii, L. and Pisani, A. and Potter, D. and Quai, S. and Radovich, M. and Rocci, P.-F. and Sacquegna, S. and Sahlén, M. and Sanders, D. B. and Schneider, A. and Sciotti, D. and Sellentin, E. and Smith, L. C. and Sorce, J. G. and Tanidis, K. and Tao, C. and Testera, G. and Teyssier, R. and Tosi, S. and Troja, A. and Tucci, M. and Valieri, C. and Venhola, A. and Vergani, D. and Verza, G. and Walton, N. A.},
	month = may,
	year = {2025},
	note = {arXiv:2505.04688 [astro-ph]},
}

@article{linder_exploring_2003,
	title = {Exploring the {Expansion} {History} of the {Universe}},
	volume = {90},
	issn = {0031-9007, 1079-7114},
	url = {http://arxiv.org/abs/astro-ph/0208512},
	doi = {10.1103/PhysRevLett.90.091301},
	number = {9},
	urldate = {2025-11-03},
	journal = {Physical Review Letters},
	author = {Linder, Eric V.},
	month = mar,
	year = {2003},
	note = {arXiv:astro-ph/0208512},
	pages = {091301},
}

\appendix

\section{BAO fitting results}\label{Apd:BAO fitting}

\begin{table}[!h]
\centering
\scalebox{0.90}{%
\setlength\tabcolsep{4pt}
\renewcommand\arraystretch{1.4}
\makebox[\textwidth][c]{
\begin{tabular}{c|c|c|c|c|c|c}
\toprule
Catalogs & clean  &  dv-LRG &  dv-QSO-G  & dv-QSO-L  & dv-ELGcatas  &  dv-slitless\\
\hline \hline
$\aiso$ & $1.003 \pm 0.005 $ & $1.003\pm0.005$ & $1.002\pm0.005$ & $1.003\pm0.004$ 
        & $1.003\pm0.005$ & $1.003\pm0.005$ \\
\hline 
$\aap$  & $1.002\pm0.016$  & $0.998\pm0.016$ &  $0.998\pm0.016$ & $0.999\pm0.017$ 
        & $1.000\pm0.017$ &  $0.998\pm0.017$  \\
\hline
$\chi^2/\rm{dof}$  & $1.102\pm0.274$  & $ 1.090\pm0.250$ &  $1.109\pm0.259$ & $1.064\pm0.229$ 
        &  $1.074\pm0.222$ &  $1.061\pm0.230$  \\
\bottomrule
\end{tabular}}
}
\caption{Comparison of BAO best-fit results for the clean and contaminated mock catalogs described in \cref{tab:mock catalogs}. Table reports the mean and standard deviation of best-fit values from 200 fits. The $\aiso$ and $\aap$ constraints remain unbiased in the presence of redshift errors. The corresponding best-fit $\chi^2$ over the degree of freedom (dof) shows the same level of goodness-of-fit.}
\label{tab:BAO constrain}
\end{table}

BAO serves as a standard ruler in cosmology and is generally considered robust against a wide range of systematic effects. In this work, we also test whether BAO measurements remain unbiased under the spectroscopic redshift errors. BAO analysis focuses on constraining two dilation parameters, $\alpha_\parallel$ and $\alpha_\perp$, which are reparameterized into two scaling parameters, $\aiso$ and $\aap$, respectively. These two parameters provide information about the isotropic BAO scale and AP effect. We adopt the BAO modeling framework implemented in DESI DR1 analysis \cite{desi_collaboration_desi_2025-2} using the \desilike~package. The results are obtained by fitting the monopole and quadrupole of the two-point correlation function from the mean of 25 randomly selected mocks out of 500 realizations without BAO reconstruction. The covariance matrix is built from 500 mocks and rescaled to an effective volume of $V_{25} = 25 \, \hgpc$.

\cref{tab:BAO constrain} summarizes the mean and standard deviation for $\aiso$ and $\aap$ best-fit values and $\chi^2$ obtained from 200 independent fits. The inferred $\aiso$ and $\aap$ values show only minor differences from the clean case, both in their means and standard deviations. The best-fit $\chi^2$ values also show no increase after including redshift errors, indicating a same level of goodness of fit. Therefore, we conclude that BAO measurements remain unbiased and robust against spectroscopic redshift errors at the current level of precision.

\section{$\chi^2$ results for baseline analysis}\label{Apd:Supplementary figures}

\begin{figure}
    \centering
    \includegraphics[width=0.8\linewidth]{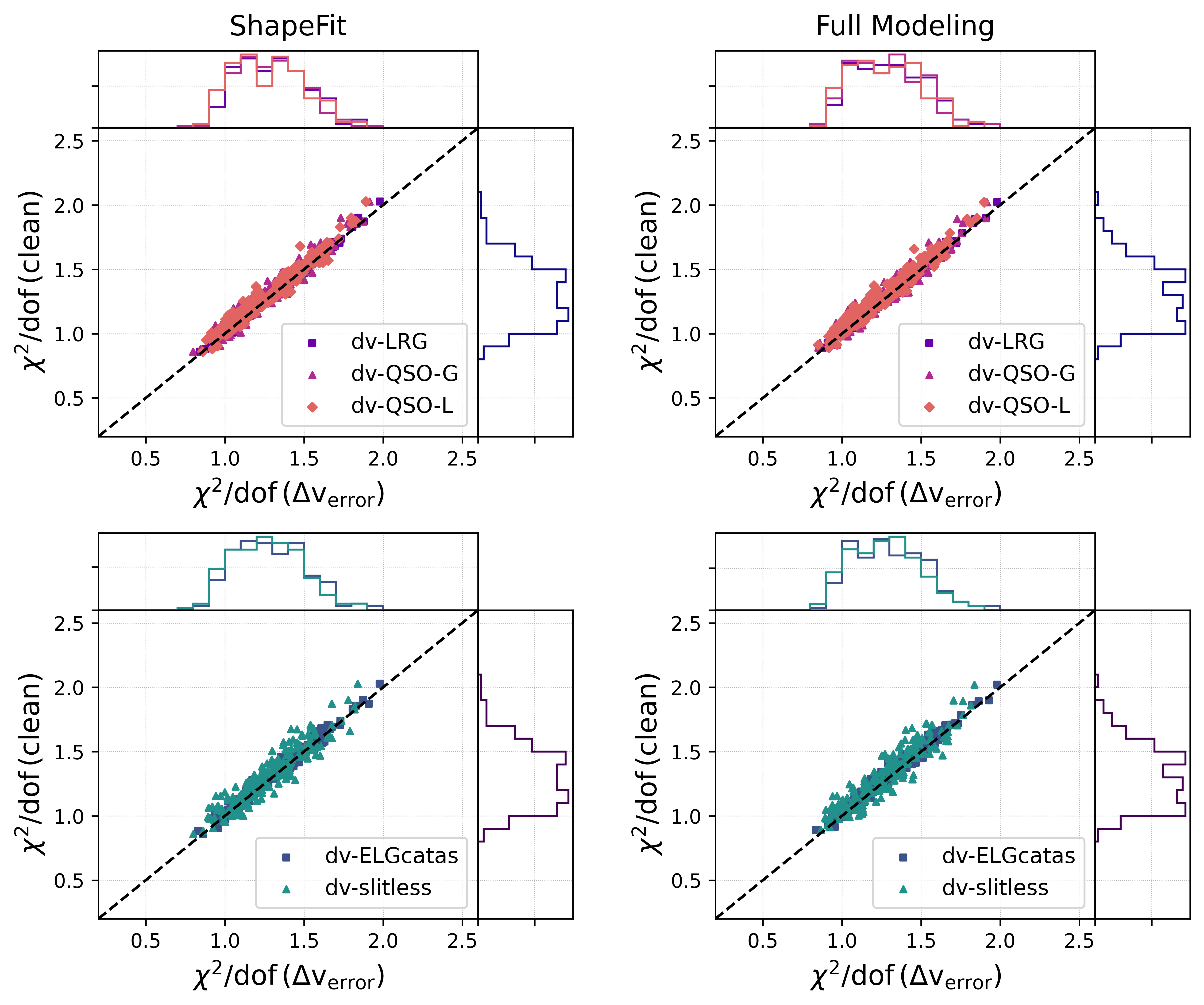}
    \caption{Comparison of corresponding $\chi^2$ for \cref{fig:scatter_SF_bestfit} and \cref{fig:scatter_FM_bestfit}. The best-fit $\chi^2$ values are normalized by the degrees of freedom, with results for ShapeFit (left panels) and Full Modeling (right panels). For each point, we fit the mean of 25 random mocks, with the covariance rescaled to $V_{25}$. The difference of $\chi^2$ is small between clean and contaminated mock catalogs, demonstrating that all fits achieve the same level of goodness.}
    \label{fig:scatter_chi2}
\end{figure}

\cref{fig:scatter_chi2} presents the distribution of best-fit $\chi^2$ values for the baseline analysis. The left two panels show results for the ShapeFit analysis, corresponding to the scatter plot in \cref{fig:scatter_SF_bestfit}, while the right two panels show the Full-Modeling results for \cref{fig:scatter_FM_bestfit}. From the $\chi^2$ distributions, we do not observe significant differences (<$15\%$) or systematic shifts between the clean and contaminated mocks analyses. This indicates that the theoretical model provides a same level of good prediction with the observational data, even potential deviations in the parameter estimates may arise under contamination.

\end{document}